\documentclass[conference]{IEEEtran}
\ifCLASSINFOpdf
  \usepackage[pdftex]{graphicx}
  \graphicspath{{../pdf/}{../jpeg/}}
  \DeclareGraphicsExtensions{.pdf,.jpeg,.png}
\else
  \usepackage[dvips]{graphicx}
  \graphicspath{{../eps/}}
  \DeclareGraphicsExtensions{.eps}
\fi
\usepackage{amsmath}
\ifCLASSOPTIONcompsoc 
\usepackage[tight,normalsize,sf,SF]{subfigure} 
\else 
\usepackage[tight,footnotesize]{subfigure} 
\fi 

\usepackage{cite}
\usepackage{url}
\usepackage{pifont}
\usepackage{flushend}

\usepackage{listings}

\begin{document}

\title{Evaluating the Performance Impact of Multiple Streams on the MIC-based Heterogeneous Platform}

\author{\IEEEauthorblockN{Zhaokui Li, Jianbin Fang, Tao Tang, Xuhao Chen, Cheng Chen and Canqun Yang}
\IEEEauthorblockA
{Software Institute, School of Computer, National University of Defense Technology, Changsha, China\\
Email: zhaokuili@yeah.net, \{j.fang, taotang84, chenxuhao, chencheng, canqun\}@nudt.edu.cn}
}

\maketitle

\begin{abstract}

Using \textit{multiple streams} can improve the overall system performance by mitigating the data transfer overhead on heterogeneous systems. Prior work focuses a lot on GPUs but little is known about the performance impact on (Intel Xeon) Phi. In this work, we apply multiple streams into six real-world applications on Phi.  We then systematically evaluate the performance benefits of using multiple streams. The evaluation work is performed at two levels: the microbenchmarking level and the real-world application level. Our experimental results at the microbenchmark level show that data transfers and kernel execution can be overlapped on Phi, while data transfers in both directions are performed in a serial manner. At the real-world application level, we show that  both overlappable and non-overlappable applications can benefit from using multiple streams (with an performance improvement of up to 24\%).  We also quantify how task granularity and resource granularity impact the overall performance. Finally, we present a set of heuristics to reduce the search space when determining a proper task granularity and resource granularity.  To conclude, our evaluation work provides lots of insights for runtime and architecture designers when using multiple streams on Phi. 

\end{abstract}
\IEEEpeerreviewmaketitle

\section{Introduction}


Heterogeneous platforms are increasingly popular in many application domains~\cite{citeulike:2767438}. The combination of using a host CPU combined with a specialized processing unit (e.g., GPGPUs or Intel Xeon Phi) has been shown in many cases to improve the performance of an application by significant amounts. Typically, the host part of a heterogeneous platform manages the execution context while the time-consuming code piece is offloaded to the coprocessor. Leveraging such platforms can not only enable the achievement of  high peak performance, but increase the \textit{performance per Watt ratio}.



Given a heterogeneous platform, how to realize its performance potentials  remains a challenging issue. In particular, programmers need to explicitly move data between host and device over \texttt{PCIe} before and/or after running kernels. The overhead counts when data transferring takes a decent amount of time, and determines whether to perform offloading is worthwhile~\cite{citeulike:13920330, citeulike:6102210, citeulike:13920339}. To hide this overhead, overlapping kernel executions with data movements is required. To this end, \textit{multiple streams} (or \textit{streaming mechanism}) has been introduced, e.g., CUDA Streams~\cite{tr:cuda:best}, OpenCL Command Queues~\cite{website:opencl_ref}, and Intel's hStreams~\cite{tr:hstreams:arch}. These implementations of \textit{multiple streams} spawn more than one streams/pipelines so that the data movement stage of one pipeline overlap the kernel execution stage of another~\footnote{In the context, the streaming mechanism is synonymous with \textit{multiple streams}, and thus we refer the \textit{streamed code} as \textit{code with multiple streams}.}.


Prior works on multiple streams mainly focus on GPUs and the potential of using multiple streams on GPUs is shown to be significant~\cite{citeulike:13920334, citeulike:9715521, citeulike:13920353, DBLP:journals/ieicet/InoNH13}.  Liu et al. give a detailed study into how to achieve optimal task partition within an analytical framework for AMD GPUs and NVIDIA GPUs~\cite{citeulike:13920353}. In~\cite{citeulike:9715521}, the authors model the performance of asynchronous data transfers of CUDA streams to determine the optimal number of streams. However, little is known about how multiple streams behave on the OS-enabled coprocessor such as Intel Xeon Phi. For such coprocessors, programmers can explicitly map streams to different groups of cores, i.e., they have control of \textit{resource granularity}. This control on GPUs is not exposed to programmers. Thus, how resource granularity impacts the overall performance and how to determine a proper resource granularity on Phi is unknown.



To answer these questions and gain insights of using multiple streams on Phi, we provide a systematic performance evaluation. Specifically, we evaluate the performance impact of multiple steams at two levels: the microbenchmarking level and the real-world application level. At the microbenchmarking level, we measure the overlapping capability of Phi with multiple streams including \ding{192} the overlapping of data transfers in both directions, \ding{193} the overlapping of data transfers with kernel execution, and \ding{194} performance potentials from resource partitioning. At the real-world application level, we first give a performance comparison of applications with and without using multiple streams, and then provide an in-depth analysis of the performance factors by using six different real-world applications. Also, we present a set of heuristics to reduce the huge search space when selecting a proper task granularity and resource granularity. Our preliminary results on multiple Phis show a significant performance improvement (over 1 Phi) without code changes and we conclude that using multiple streams is a promising programming tool for multiple devices. 







To summarize, our contributions are as follows. 

\begin{itemize}

\item We apply the multi-stream mechanism to 7 (6+1) applications representative of different domains and patterns. With them, we systematically evaluate the capability of multiple streams on Phi.


\item We quantify how each performance factor impacts the application performance and perform an in-depth analysis on the performance changes. 

\item We present a set of guidelines to significantly reduce the search space when determining task granularity and/or resource granularity, based on our observations.


\item We give a preliminary performance evaluation on multiple MICs with multiple streams. 

\end{itemize}

%

The rest of this paper is organized as follows: Section~\ref{sec:hstr_background} gives a brief description of multiple streams in terms of temporal resource sharing and spatial resource sharing, a prototype implementation of multiple streams, and the related work. Section~\ref{sec:hstr_setup} introduces the experimental setup and the benchmarks. We provide a systematic performance evaluation of multiple streams with microbenchmarks in Section~\ref{sec:microbench} and with real-world applications in Section~\ref{sec:overall}. Section~\ref{sec:multimics} discusses the performance of multiple streams on multiple devices and Section~\ref{sec:conclusion} concludes the work. 



\section{Background} \label{sec:hstr_background}
In this section, we introduce multiple streams in terms of temporal sharing and spatial sharing, describe a multiple stream prototype (\texttt{hStreams}), and give the related work. 

\subsection{Multiple Streams} \label{subsec:multiple:streams}
\subsubsection{Temporal Sharing}
Using multiple streams can overlapping computation and communication (data transfers), thus realizing the \textit{temporal sharing} of resources. On the whole, we divide the offloading process of heterogeneous applications into three steps: (1) move data from host to device (\texttt{H2D}), (2) kernel execution (\texttt{EXE}), and (3) move data from device to host (\texttt{D2H}).  Overlapping these three steps will significantly improve the overall performance. 

\begin{figure}[!h]
\centering
\includegraphics[width=0.46\textwidth]{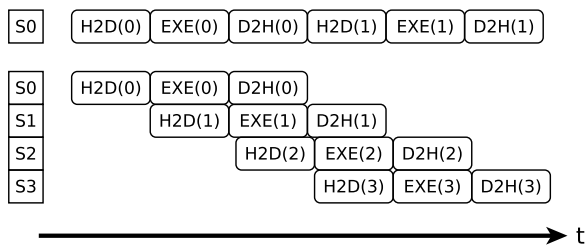}
\caption{Temporal sharing.}
\label{fig:temporal_sharing}
\end{figure}

Figure~\ref{fig:temporal_sharing} shows an illustrative comparison between single stream and multiple streams. Suppose that the aforementioned three steps consume equal amount of time for a given task. When using a single stream (i.e., the steps run in a serial manner), the code takes 6 time units to finish two tasks. Meanwhile,  the multiple-streamed code will finish four tasks (using four streams) in the same amount of time. 

\subsubsection{Spatial Sharing}
\begin{figure}[!h]
\centering
\includegraphics[width=0.40\textwidth]{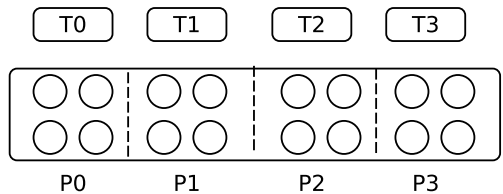}
\caption{Spatial sharing. The circles represent processing cores, \texttt{Tx} represents a task, and \texttt{Px} represents a partition. }
\label{fig:spatial_sharing}
\end{figure}

Using multiple streams also enjoys the idea of resource partitioning. That is, to partition the resource into multiple groups and map each stream onto a partition. Therefore, different streams can run on different partitions simultaneously, i.e., resource \textit{spatial sharing}. Nowadays accelerators have a large amount of processing units that some applications cannot efficiently exploit them for a given task. Typically, we offload a task and let it occupy all the processing cores. Alternatively, we divide the processing cores into multiple groups (each group is named as \textit{partition}). Figure~\ref{fig:spatial_sharing} shows that a device has 16 cores and is logically divided into four partitions (\texttt{P0}, \texttt{P1}, \texttt{P2}, \texttt{P3}). Then different tasks are offloaded onto different partitions, e.g., \texttt{T0}, \texttt{T1}, \texttt{T2}, \texttt{T3} runs on \texttt{P0}, \texttt{P1}, \texttt{P2}, \texttt{P3}, respectively. In this way, we aim to improve the device utilization.  
 
\subsection{hStreams}
\begin{figure}[!h]
\centering
\includegraphics[width=0.50\textwidth]{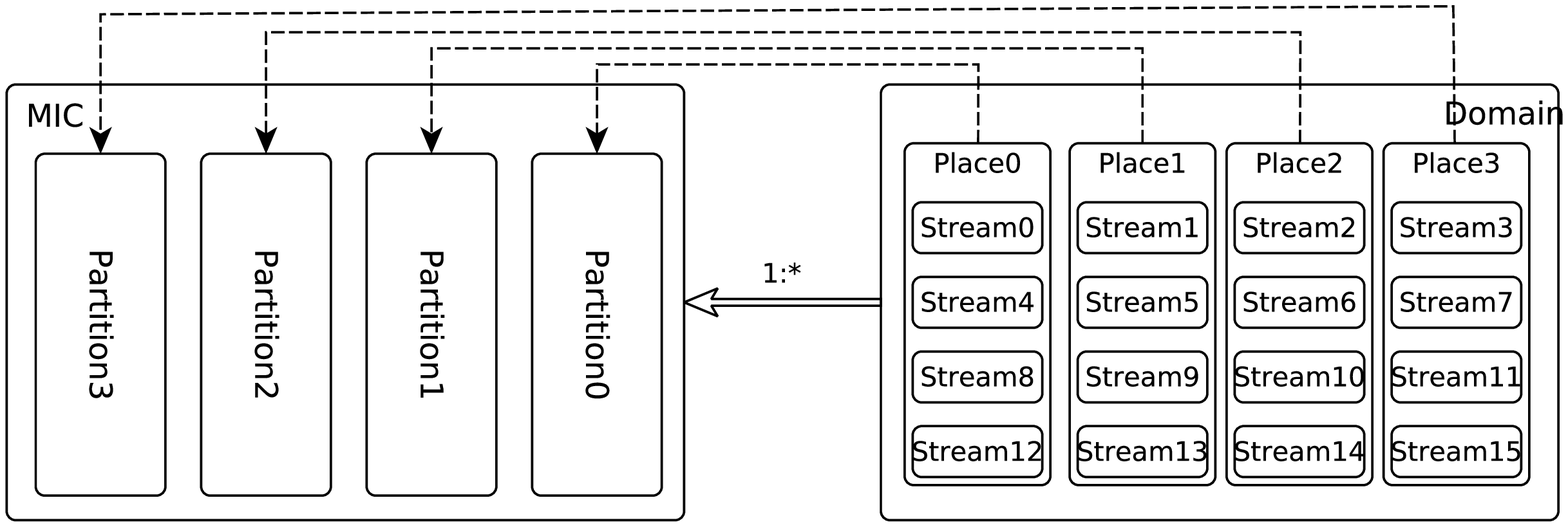}
\caption{hStreams resource view.}
\label{fig:hstreams_resource_view}
\end{figure}

\texttt{hStreams} is an open-source implementation of multiple streams from Intel. At its core is the resource partitioning mechanism~\cite{tr:hstreams:arch}. Figure~\ref{fig:hstreams_resource_view} shows the mapping between logical concepts and the physical machines (e.g., Intel Xeon Phi). At the physical level, the whole device is partitioned into multiple groups and thus each group has several processing cores. At the logical level, a device can be seen as one or more \textit{domains}. Each domain contains multiple \textit{places}, each of which then has multiples \textit{streams}. The logical concepts are visible to programmers, while the physical ones are transparent to them and the mapping between them are automatically handled by~\texttt{hStreams}.  

\texttt{hStreams} is implemented as a library and provides users with APIs to access coprocessors/accelerators efficiently. Programming with \texttt{hStreams} resembles that in \texttt{CUDA} or \texttt{OpenCL}. Programmers have to create the streaming context, move data between host and device, and invoke kernel execution. And they also have to split tasks to use multiple streams. Further, \texttt{hStreams} cannot be used alone, but has to used with other programming models such as \texttt{OpenMP} or Intel \texttt{TBB}. 

\subsection{Related Work} \label{sec:related_work}

\textbf{Pipelinining}  is widely used in modern computer architectures~\cite{citeulike:1102804}. Specifically, the pipeline stages of an instruction run on different functional units, e.g., arithmetic units or data loading units. In this way, the stages from different instructions can occupy the same functional unit in different time steps, thus improving the overall system throughput. Likewise, the heterogeneous applications are divided into stages (\texttt{H2D}, \texttt{EXE}, \texttt{D2H}), and can exploit the idea of software pipelining on the heterogeneous platforms (as mentioned in Section~\ref{subsec:multiple:streams}). 


\textbf{Multi-tasking} provides concurrent execution of multiple applications on a single device. In~\cite{citeulike:11069829}, the authors propose and make the case for a GPU multitasking technique called \textit{spatial multitasking}. The experimental results show that the proposed spatial multitasking can obtain a higher performance over cooperative multitasking. In~\cite{WendeSteinkeCordes2014}, Wende et al. investigate the concurrent kernel execution mechanism that enables multiple small kernels to run concurrently on the Kepler GPUs. Also, the authors evaluate the Xeon Phi offload models with multi-threaded and multi-process host applications with concurrent coprocessor offloading~\cite{citeulike:13920403}. Both multitasking and multiple streams share the idea of spatial resource sharing. Different from multi-tasking, using multiple streams needs to partition the workload of a single application (rather than multiple applications) into many tasks. 



\textbf{Workload Partition: }  There is a large body of workload partitioning techniques, which intelligently partition the workload between a CPU and a coprocessor at the level of algorithm~\cite{citeulike:13920419}\cite{citeulike:13920423} or during program execution~\cite{citeulike:13920424}\cite{citeulike:10176926}. Partitioning workloads aims to use unique architectural strength of processing units and improve resource utilization~\cite{citeulike:13920415}. In this work, we focus on how to efficiently utilize the coprocessing device with multiple streams. Ultimately, we need to leverage both workload partitioning and multiple streams to minimize the end-to-end execution time. 

\textbf{Multiple Streams Modeling:} In~\cite{citeulike:9715521}, Gomez-Luna et al. present performance models for asynchronous data transfers on different GPU architectures. The models permit programmers to estimate the optimal number of streams in which the computation on the GPU should be broken up. In~\cite{citeulike:13920334}, Werkhoven et al. present an analytical performance model to indicate when to apply which overlapping method on GPUs. The evaluation results show that the performance model are capable of correctly classifying the relative performance of the different implementations. In~\cite{citeulike:13920353}, Liu et al. carry out a systematic investigation into task partitioning to achieve maximum performance gain for AMD and NVIDIA GPUs. Unlike these works, we discuss the heuristics of reducing the search space when determining the factors. Using a model on Phi will be investigated as our future work.

\section{Experimental Methodology } \label{sec:hstr_setup}
In this section, we first introduce the hardware and software environment, and then describe the used benchmarks. 


\subsection{Platform Configurations}
The heterogeneous platform used in this work includes a dual-socket Intel Xeon CPU (12 cores for each socket) and an Intel Xeon 31SP Phi (57 cores for each card). The host CPUs and the cards are connected by a \texttt{PCIe} connection. As for the software, the host CPU runs Redhat Linux v6.4 (the kernel version is 2.6.32-279.el6.x86\_64), while the coprocessor runs a customized uOS (v2.6.38.8). Intel's MPSS (v3.5.2) is used as the driver and the communication backbone between the host and the coprocessor.  Also, we use Intel's multi-stream implementation~ \texttt{hStreams (v3.5.2)}. 

\subsection{Benchmarks} \label{subsec:benchmarks}
We use seven benchmarks, among which, \texttt{Matrix Multiplication} and \texttt{Cholesky Factorization} are from the hStreams SDK while the others are written by ourselves. \texttt{hBench} is used to quantify the capability of multiple steams in terms of temporal sharing and spatial sharing. \texttt{Kmeans}, \texttt{Hotspot}, \texttt{NN}, and \texttt{SRAD} from the Rodinia benchmark suite are used in our work~\cite{citeulike:6102210}. When porting them in hStreams, we partition the whole dataset into tiles, each of which represents a task. Then at least one task is mapped to a stream. The OpenMP regions are coded as kernels (like CUDA or OpenCL), which will be offloaded onto the coprocessor side. We run each benchmark for 11 iterations, ignore the first iteration, and calculate the mean results. 

Besides the benchmark descriptions, we characterize how each benchmark runs on a heterogeneous platform (e.g., the execution flow in Figure~\ref{fig:flow}). In the figure, \texttt{H2D} represents a stage of moving a block of data elements from host to device, and \texttt{D2H} means the vice verse. \texttt{EXE} represents a kernel execution stage. The arrow connection represents the dependency relationship between two continuous stages, and \texttt{sync/async} marks whether the connected stages can run asynchronously. That is, \texttt{async} means that the two connected stages can work concurrently on different data blocks. 

 
\subsubsection{hBench}
The basic operation of the microbenchmark is $B[i]=A[i]+\alpha$. Before kernel execution, we need to move array $A$ to the coprocessor, and move the output array $B$ back once the kernel finishes execution. The compute complexity of the kernel execution is controlled by $iterations$. Simply, more iterations consume more computational time. The data movements and kernel execution can be either synchronous or asynchronous. With this microbenchmark, we aim to evaluate the capability of multiple streams on the Phi coprocessor. 

\subsubsection{Matrix Multiplication (MM)}
It is an expensive mathematical operation used as a building block in many scientific applications. In the context, we divide the matrix into tiles. The execution flow of the application is shown in Figure~\ref{fig:flow:mm}. 

\subsubsection{Cholesky Factorization (CF)}
It is a decomposition of Hermitian, positive-definite matrix into the product of a lower triangular matrix and its conjugate transpose, usefull for efficient numerical solutions and Monte Carlo simulations. When it is applicable, the Cholesky factorization is roughly twice as efficient as LU factorization for solving system of linear equations. The execution flow of the application is shown in Figure~\ref{fig:flow:cf}. 

\subsubsection{Kmeans}
Clustering is a form of unsupervised learning whereby a set of observations (i.e., data points) are partitioned into natural groupings or clusters according some similarity measures. Kmeans is a clustering algorithm used to extensively in data mining. It identifies related points by associating each data point with its nearest center, computing new cluster centroids, and iterating until convergence. The initial version of the benchmark is taken from the Northwestern MineBench~\cite{citeulike:3440815}. The execution flow of this benchmark is shown in Figure~\ref{fig:flow:km}. 

\subsubsection{Hotspot}
Based on an architectural floor plan and power measurement, Hotspot is used to estimate processor temperature. Specifically, the benchmark has a 2D transient thermal simulation kernel, which solves a suite of differential equations iteratively for block temperatures. It takes power and initial temperature values of the corresponding area of the target chip as input. The flow of this benchmark is shown in Figure~\ref{fig:flow:hs}. 

\subsubsection{NN}
Nearest Neighbor finds the k-nearest neighbors from an unstructured data set. The sequential NN algorithm reads in one record at a time, caculates the Euclidean  distance from the target latitude and longitude, and evaluates the k nearest neighbors. The parallel versions read in many records at a time, execute the distance calculation on multiple threads, and the master thread updates the list of nearest neighbors. The flow of this benchmark is shown in Figure~\ref{fig:flow:nn}, which is the same as that of \texttt{MM}. 

\subsubsection{SRAD}
\begin{figure}[!h]
\centering
\subfigure[Matrix Multiplication.]{\label{fig:flow:mm}\includegraphics[width=0.32\textwidth]{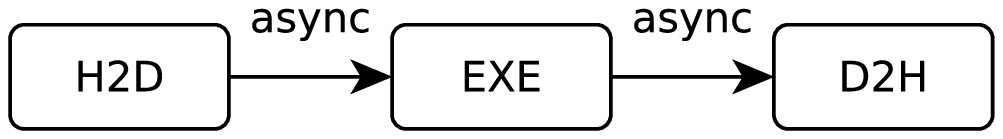}}
\subfigure[Cholesky Factorization.]{\label{fig:flow:cf}\includegraphics[width=0.45\textwidth]{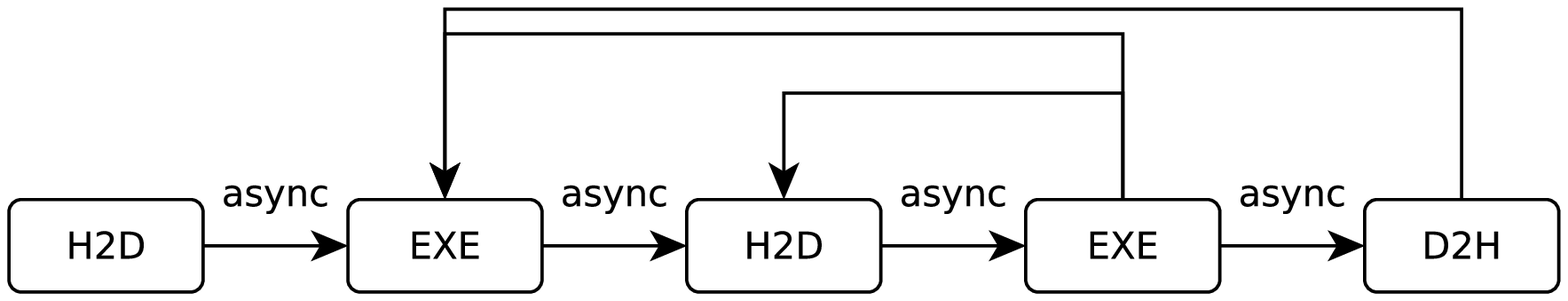}}
\subfigure[Hotspot.]{\label{fig:flow:hs}\includegraphics[width=0.32\textwidth]{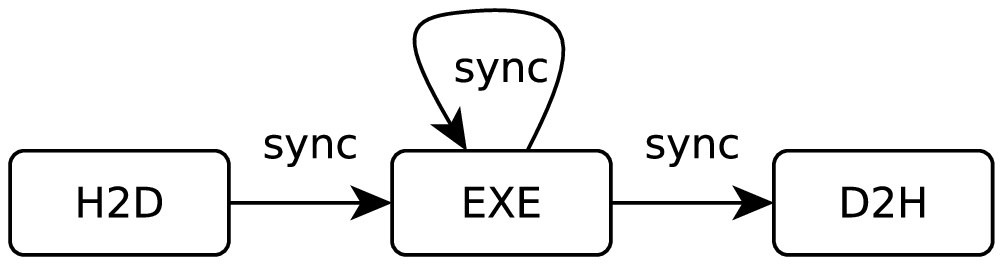}}
\subfigure[Kmeans.]{\label{fig:flow:km}\includegraphics[width=0.45\textwidth]{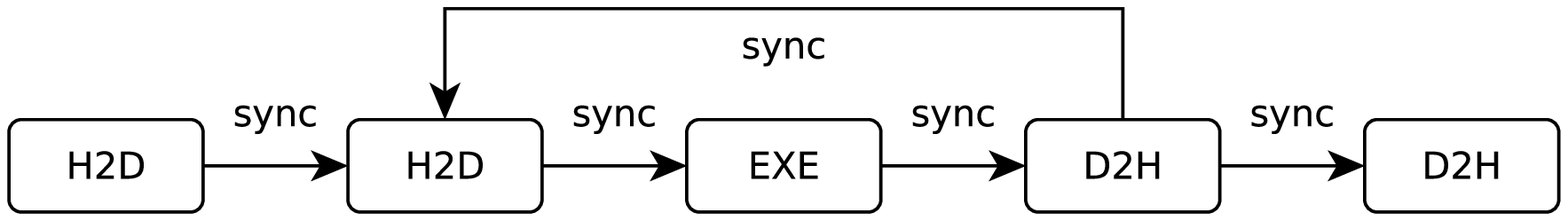}}
\subfigure[Nearest Neighbor.]{\label{fig:flow:nn}\includegraphics[width=0.32\textwidth]{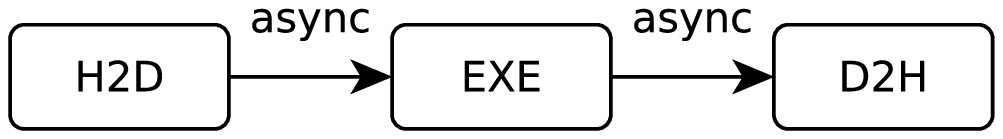}}
\subfigure[SRAD.]{\label{fig:flow:srad}\includegraphics[width=0.32\textwidth]{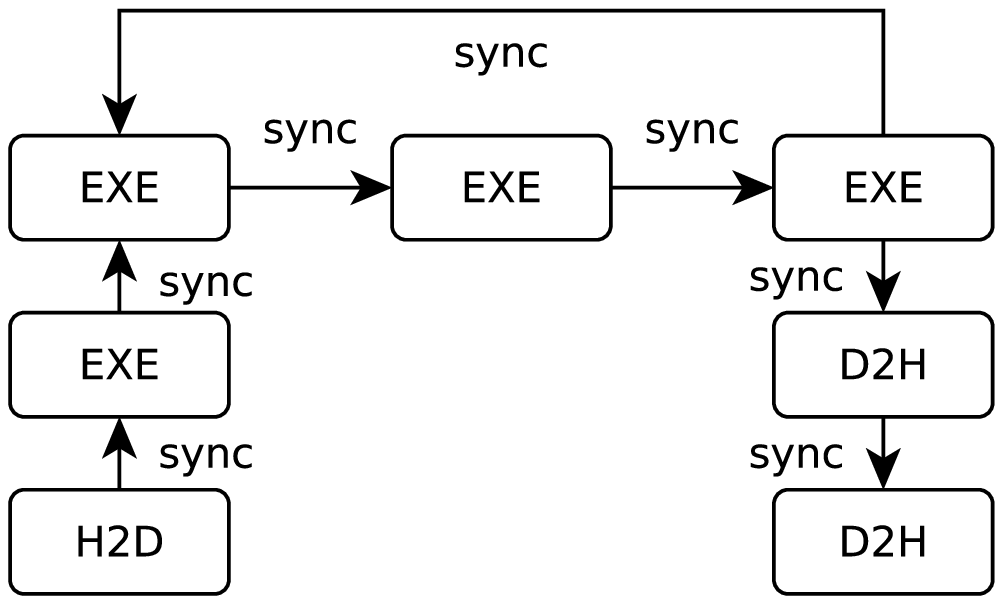}}
\caption{The flow illustration for each benchmarks. }
\label{fig:flow}
\end{figure}

Speckle Reducing Anisotropic Diffusion is a diffusion algorithm based on partial differential equations and used for removing the speckles in an image without sacrificing important image features. SRAD is widely used in ultrasonic and radar imaging applications. SRAD consists of several pieces of work: image extraction, continuous iteration (preparation, reduction, statistics, and computation), and image compression. The inputs to the program are ultrasound images and the value of each point in the computation domain depends on its four neighbors. The flow of SRAD is shown in Figure~\ref{fig:flow:srad}. 

\section{Evaluating Multiple Streams with MicroBenchmarks} \label{sec:microbench}
In this section, we evaluate the performance impact of using multiple streams from the perspective of temporal sharing and spatial sharing. First, we will evaluate the overlapping of data transfers and computations, and between different data transfers. Then, we will quantify the performance impact of using the resource partitioning (i.e., spatial sharing).

\subsection{Temporal Sharing}

\subsubsection{Overlapping Data Transfers}
Data transfers can be performed from host to device and vice verse. However, whether data can be transferred from both directions concurrently depends on the target platform. Thus, we use \texttt{hBench} to measure the overlapping of data transfers. Specifically, it moves $hd$ blocks of data elements from host to device, and moves $dh$ blocks of data elements from device to host. 

\begin{figure}[!h]
\centering
\includegraphics[width=0.40\textwidth]{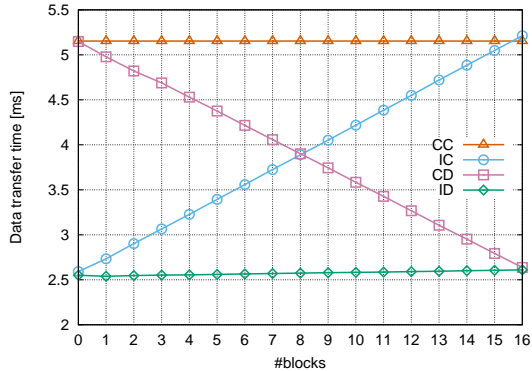}
\caption{How the data transfer time over the number of transferred blocks. }
\label{fig:ubench:dt}
\end{figure}

Figure~\ref{fig:ubench:dt} shows four cases when $dh$ and $hd$ are assigned with different values, and the block size is $1$ MB.  For \texttt{CC}, $hd=dh=16$. We first transfer 16 data blocks from host to device, and then transfer another 16 blocks from device to host. Since the total amount of data blocks remains constant, the data transfer time does not change (5.2 ms).  For \texttt{IC}, $hd$ increases from 0 to 16 and $dh=16$, while for \texttt{CD}, $hd=16$ and $dh$ decreases from 16 to 0. We observe that the data transfer time increases linearly over blocks for $IC$, while it decreases linearly for $CD$. For \texttt{ID}, $hd$ increases from 0 to 16, $dh$ decreases from 16 to 0, and $hd+dh=16$. In this case, the total amount of transferred data keeps constant, and we notice the transfer time also remains around 2.5 ms. If moving data from host to device can overlap the data transfers in the other direction, the total transferring time will be dominated by the one with more data blocks, other than the sum of transferring time. Therefore, we conclude that data transfers from both directions are performed in a serial manner. 

\subsubsection{Overlapping Data Transfers with Computation}
We then measure the overlapping of data transfers and computation with \texttt{hBench}. Apart from data transfers, we also measure the kernel execution time. The kernel computes $B[i]=A[i]+\alpha$, where array $A$ is transferred from host to device before kernel execution while array $B$ is transferred from device to host as the output. We control the computation amount by iterating the addition operation while keeping the size of array $A$ and array $B$ fixed (16MB). 


\begin{figure}[!h]
\centering
\includegraphics[width=0.40\textwidth]{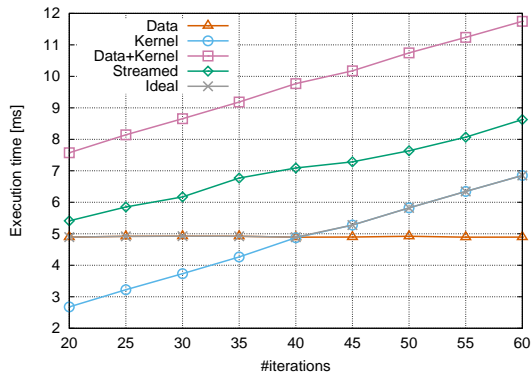}
\caption{The overlapping extent of data transfers and computation when changing the number of kernel iterations. }
\label{fig:ubench:comp}
\end{figure}

Figure~\ref{fig:ubench:comp} shows the overlapping of data transfers and computation. $Data$ (the line with triangular marks) represents the data transferring time from both directions, and $Kernel$ (the line with circle marks) represents the kernel execution time. Since the size of array $A$ and $B$ is constant, the total transferring time does not change with the number of kernel iterations. By contrast, the kernel execution time increases linearly as shown in Figure~\ref{fig:ubench:comp}. These two lines intersect when using 40 iterations in the kernel. When using less than 40 iterations, the performance is dominated by data transfers (i.e., \textit{dominant transfers}), and when using over 40 iterations, the performance is dominated by kernel execution (i.e., \textit{dominant kernel})~\cite{citeulike:9715521}. We expect that the line with cross marks in Figure~\ref{fig:ubench:comp}  represents a full overlap of data transfers and kernel execution. However, the measured execution time (the line with diamond marks) is longer than the expected execution time, illustrating the difficulty of achieving a full overlap. 



\subsection{Spatial Sharing} \label{subsec:ubench:kernel}
\begin{figure}[!h]
\centering
\includegraphics[width=0.40\textwidth]{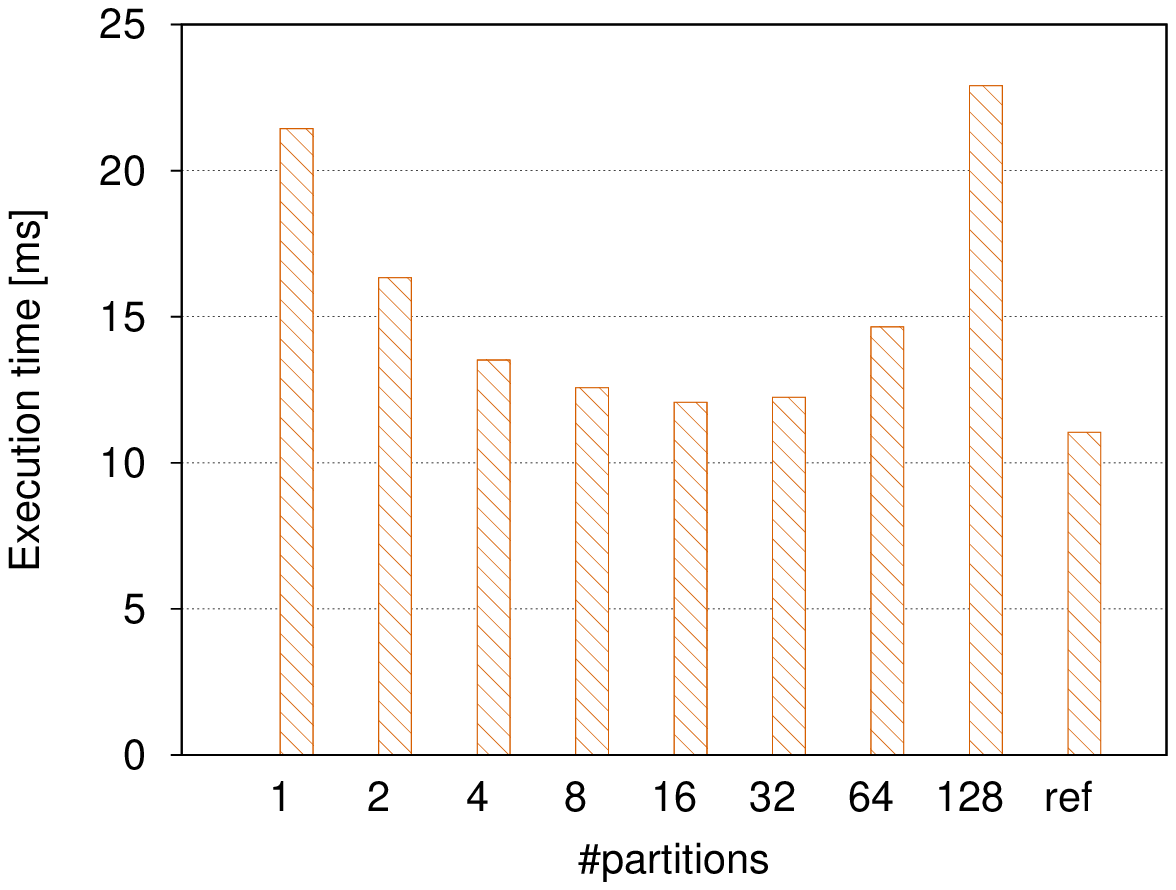}
\caption{How resource granularity impacts  the overall performance. The \texttt{ref} bar represents the execution time of the non-streamed non-tiled code. }
\label{fig:ubench:kernel}
\end{figure}

We further use \texttt{hBench} to evaluate the performance impact of resource granularity. Specifically, we partition array $A$ and $B$ into 128 blocks, and use 100 kernel iterations. Figure~\ref{fig:ubench:kernel} shows how the kernel execution (excluding the data transferring time) changes over resource granularity. We observe that, for a given task granularity,  the overall performance first increases and then decreases when changing the number of partitions. This is because partition resources can lead to a better utilization. Meanwhile, using more partitions (and streams) also introduces extra management overheads. 

The \texttt{ref} bar of Figure~\ref{fig:ubench:kernel} shows the execution time of the non-streamed non-tiled code. We see that it is lower than that of the tiled streamed code. In other words, simply partitioning the hardware resources brings no performance improvement for the kernel execution. The root reason is that we explicitly make a synchronization between data transfers and kernel execution, and the application is \textbf{\textit{non-overlappable}}. Our experimental results in Figure~\ref{fig:ubench:comp} and Figure~\ref{fig:ubench:kernel} show that using multiple streams is beneficial only when the target application is \textbf{\textit{overlappable}}. Among the aforementioned applications in Section~\ref{subsec:benchmarks} and Figure~\ref{fig:flow},  \texttt{MM}, \texttt{CF}, and \texttt{NN} are overlappable, while \texttt{Hotspot}, \texttt{Kmeans}, and \texttt{SRAD} are non-overlappable. Therefore, we expect that the first three applications can benefit from temporal and spatial sharing of hardware resources. 

\section{Evaluating Multiple Streams with Real-world Applications} \label{sec:overall}
We use 6 real-world applications to evaluate the performance impact of multiple streams. We first give an overall performance comparison of non-streamed code and streamed code. Then we analyze the performance gaps between them, and how each factor impacts the application performance. 

\subsection{An Overall Performance Comparison}

In this section, we will give an overall comparison between the streamed code (\texttt{w/}) and the non-streamed version (\texttt{w/o}).  Note that the non-streamed version uses a single stream and a single tile/task. Meanwhile, the whole datasets of streamed code are partitioned into a large number of tasks, where the task granularity is controlled by the number of tiles. Then the tasks are mapped to streams and each stream runs multiple tasks. At the low level, the hardware resources (i.e., processing cores) are partitioned into group, and the number of processing cores per partition is referred to be as resource granularity. In the experiments, we empirically enumerate all the possible values of task granularity and resource granularity to obtain the optimal performance. 



\begin{figure*}[!t]
\centering
\subfigure[MM.]{\label{fig:overall:mm}\includegraphics[width=0.32\textwidth]{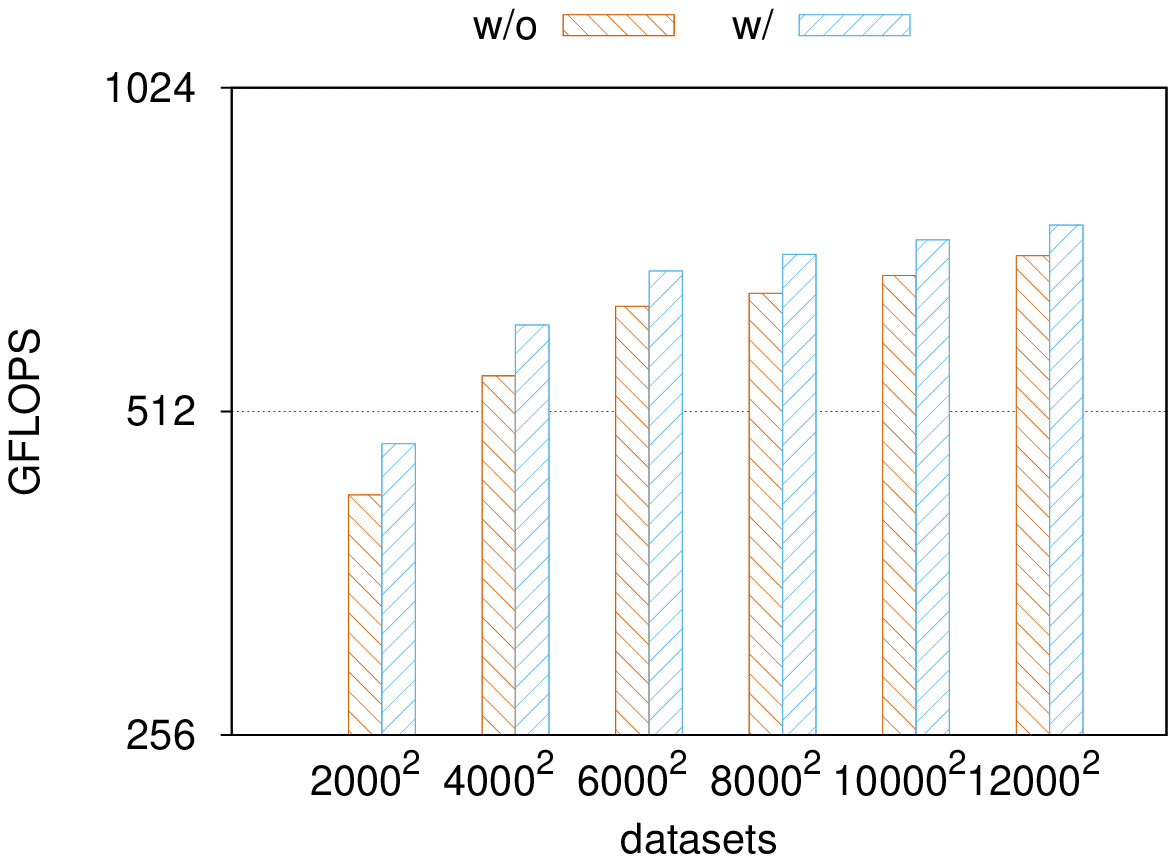}}
\subfigure[CF.]{\label{fig:overall:cf}\includegraphics[width=0.32\textwidth]{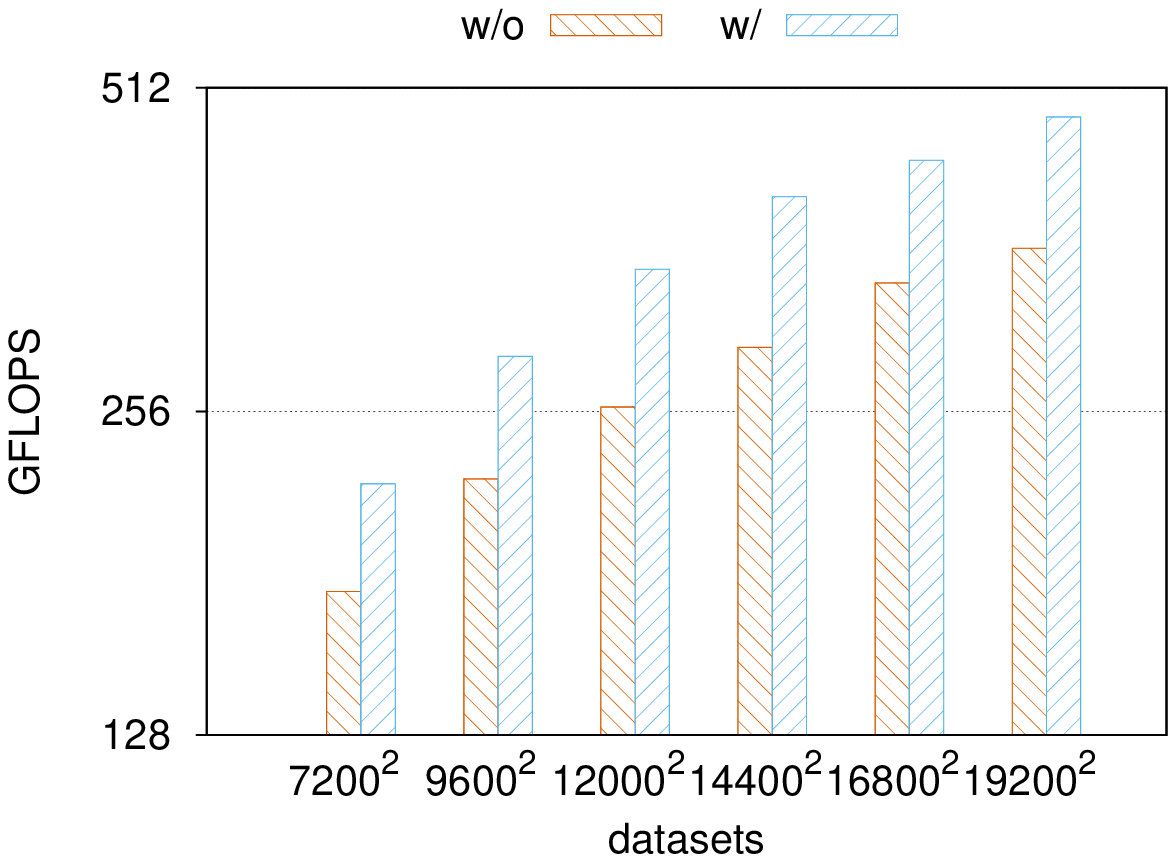}}
\subfigure[Kmeans.]{\label{fig:overall:km}\includegraphics[width=0.32\textwidth]{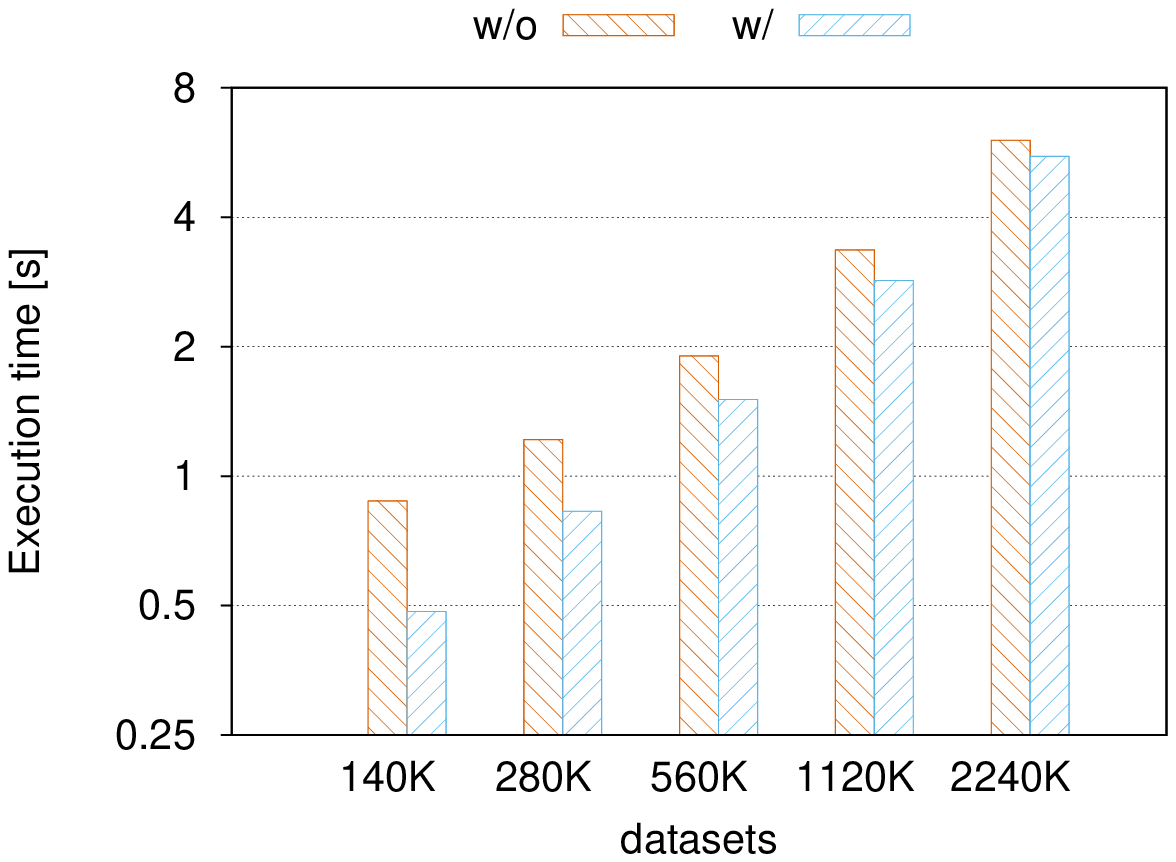}}
\subfigure[Hotspot.]{\label{fig:overall:mm}\includegraphics[width=0.32\textwidth]{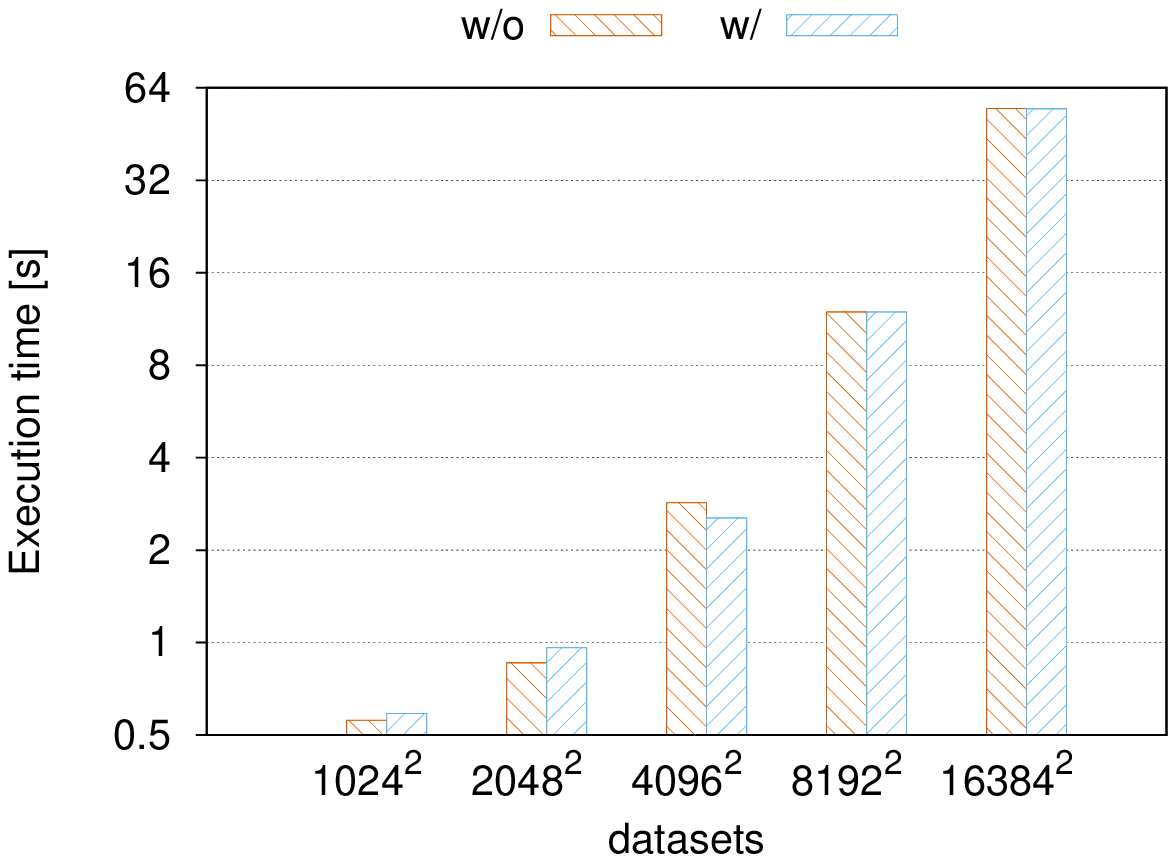}}
\subfigure[NN.]{\label{fig:overall:nn}\includegraphics[width=0.32\textwidth]{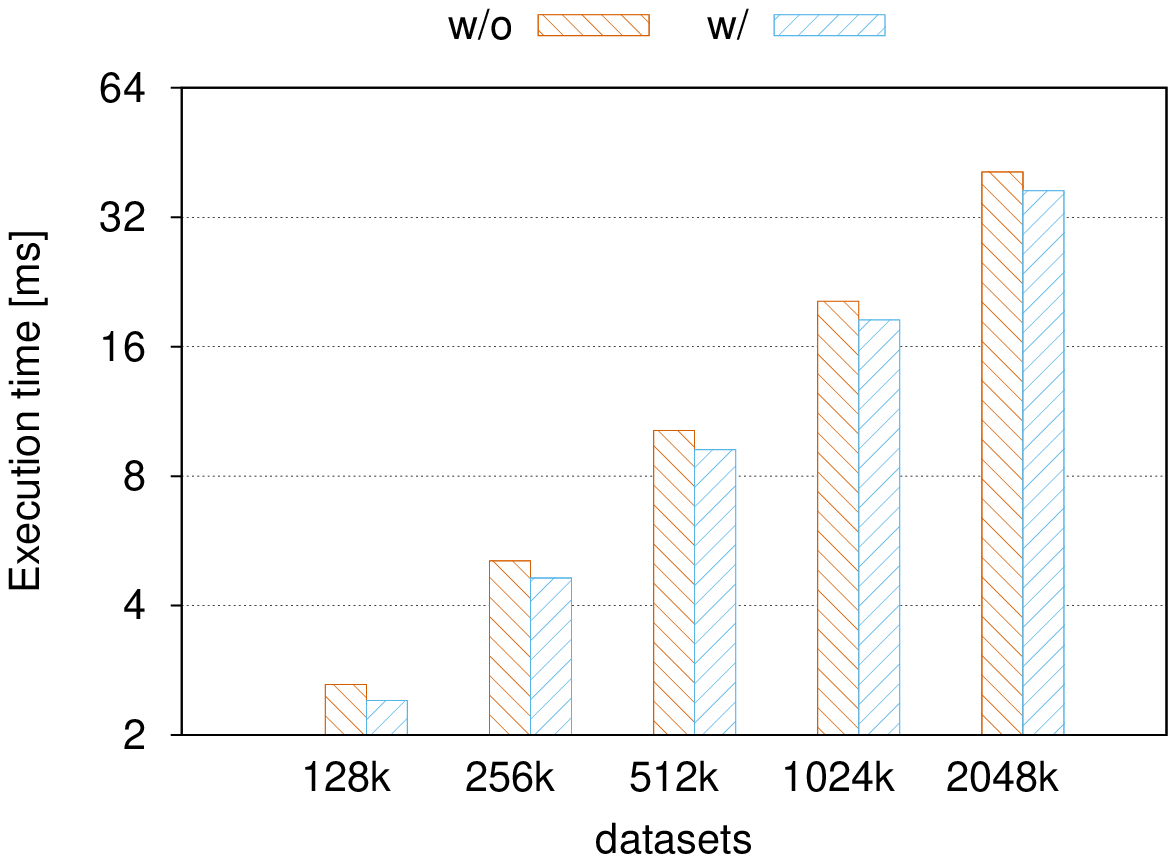}}
\subfigure[SRAD.]{\label{fig:overall:srad}\includegraphics[width=0.32\textwidth]{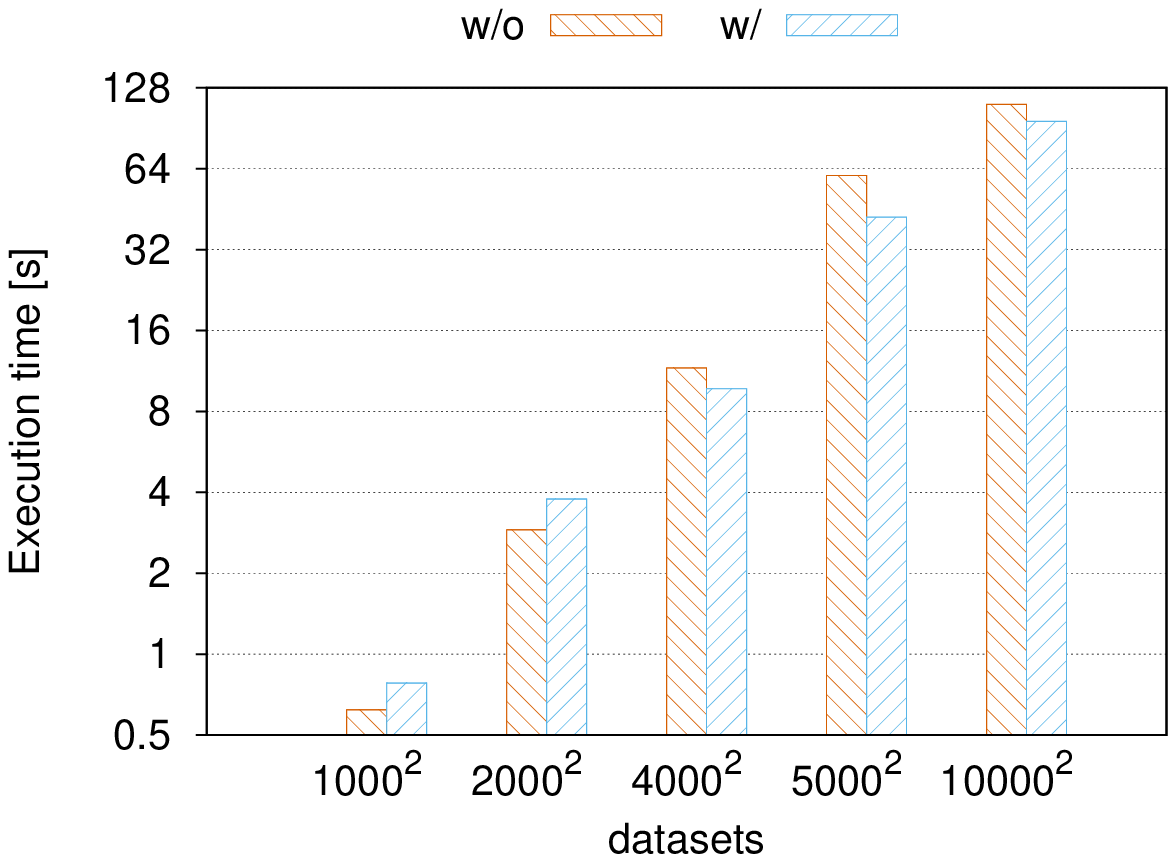}}
\caption{A performance comparison between using a single stream and multiple streams. For Kmeans, the number of centroid is 8, and we run 100 iterations before reaching convergence. For Hotspot, we run 50 simulation iterations. For NN, the target coordination is (40, 120), and the number of nearest neighbors to find is 10. For SRAD, $\lambda=0.5$, and we run the kernel for 100 iterations.}
\label{fig:overall}
\end{figure*}

Figure~\ref{fig:overall} shows the overall performance comparison. For \texttt{MM}, \texttt{CF}, \texttt{Kmeans}, and \texttt{NN}, we see that the streamed code outperforms the non-streamed code for all the used datasets, with an average performance improvement of $8.3\%$, $24.1\%$, $24.1\%$,  and $9.2\%$,  respectively.  Among the four applications, three (\texttt{MM}, \texttt{CF}, \texttt{NN}) can overlap the data transfer stage and the kernel execution stage, i.e., they are overlappable. Although \texttt{Kmeans} is an non-overlappable application, it can benefit from the reduced memory allocation and deallocation during kernel execution by employing multiple streams. 

Moreover, we see that using multiple streams brings no performance change for \texttt{Hotspot}. This is because data transfers and kernel execution of this application cannot be overlapped. Partitioning a large workload into several small workloads which are then mapped onto different resource partitions, gives no performance boost. Also, due to the overheads of managing streams, we notice that the streamed code runs slightly slower than the non-streamed code for the small datasets. 

For \texttt{SRAD}, the streamed code runs slower than the non-streamed code for small datasets. The reason resembles that of the \texttt{Hotspot}. While the streamed version of \texttt{SRAD} outperforms the non-streamed one for large datasets. This case is out of our expectation. Theoretically, it should not occur due to its non-overlappable feature. The reason is still under investigation. 


\subsection{Performance Analysis} \label{subsec:perf:analysis}

\subsubsection{How the number of partitions impacts performance?} \label{subsec:partitions}
\begin{figure*}[!t]
\centering
\subfigure[MM.]{\label{fig:partitions:mm}\includegraphics[width=0.32\textwidth]{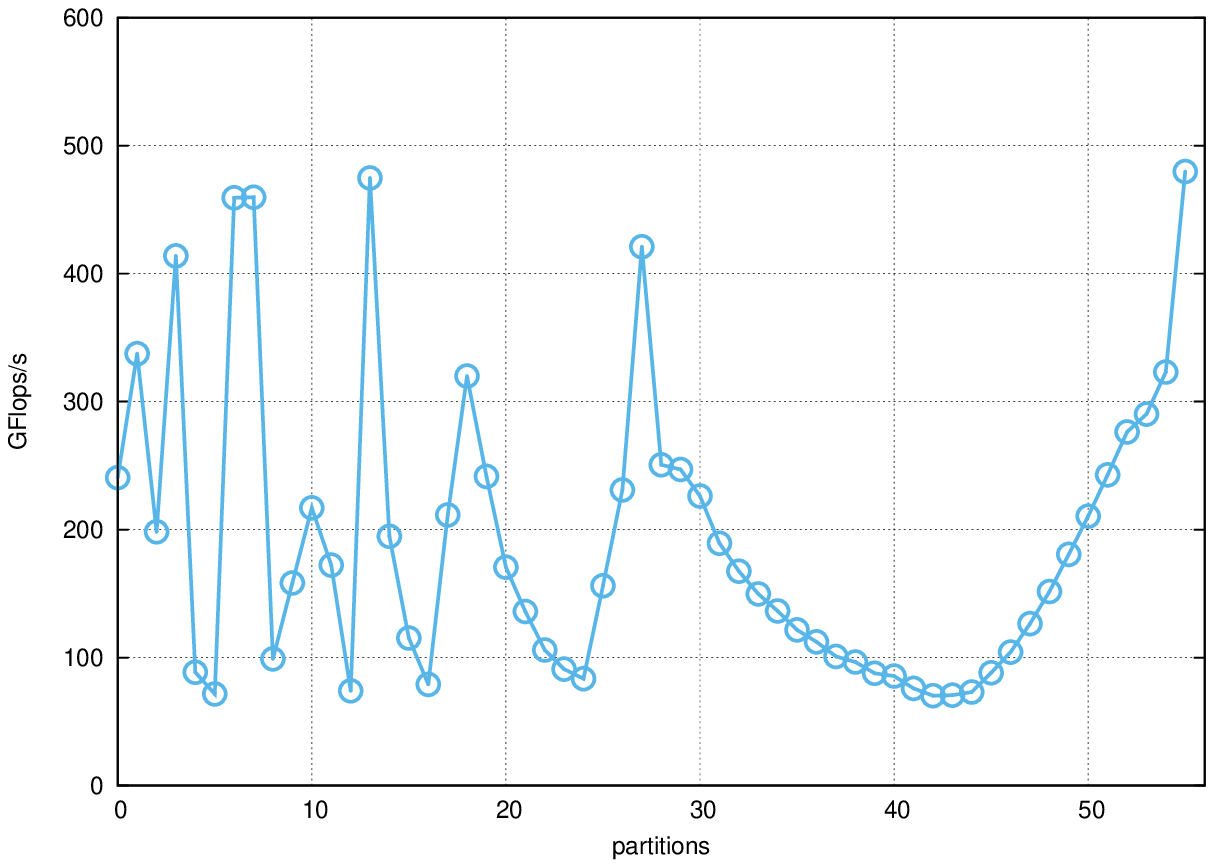}}
\subfigure[CF.]{\label{fig:partitions:cf}\includegraphics[width=0.32\textwidth]{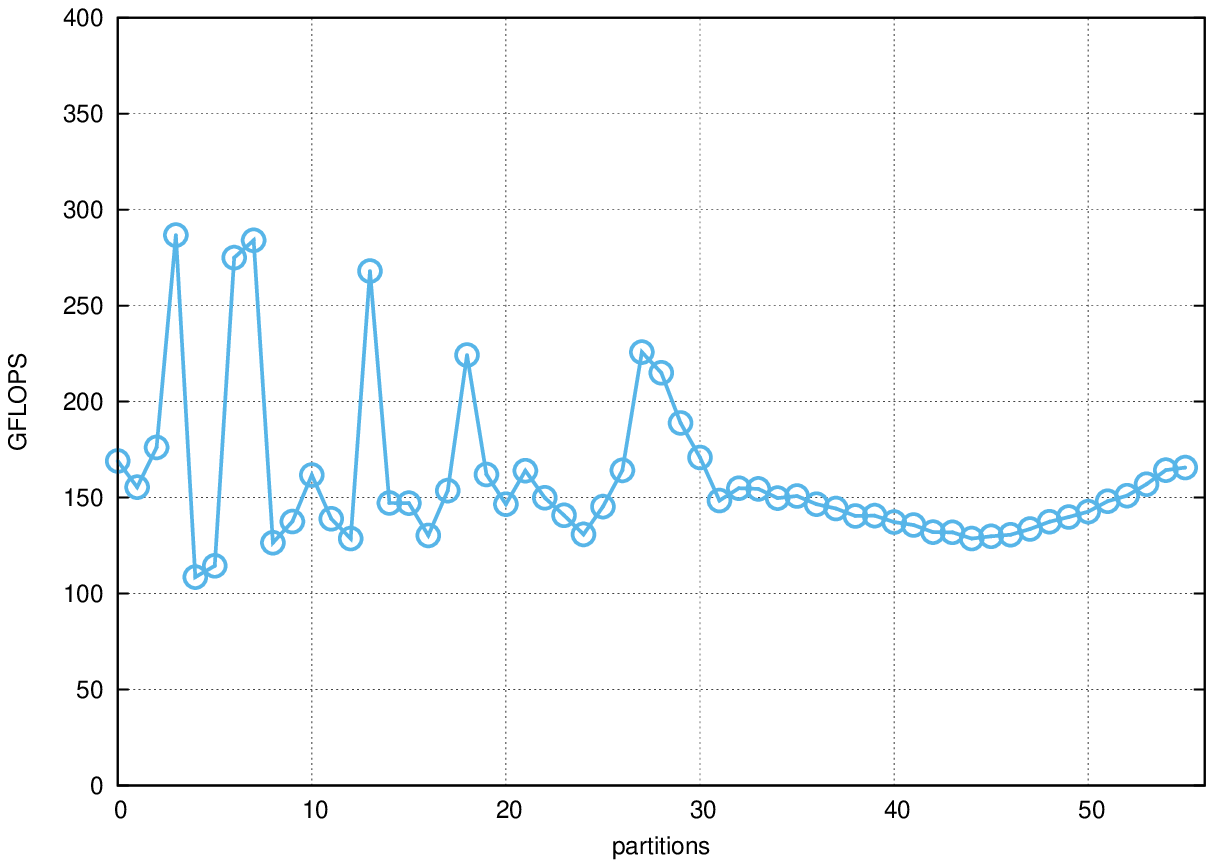}}
\subfigure[Kmeans.]{\label{fig:partitions:km}\includegraphics[width=0.32\textwidth]{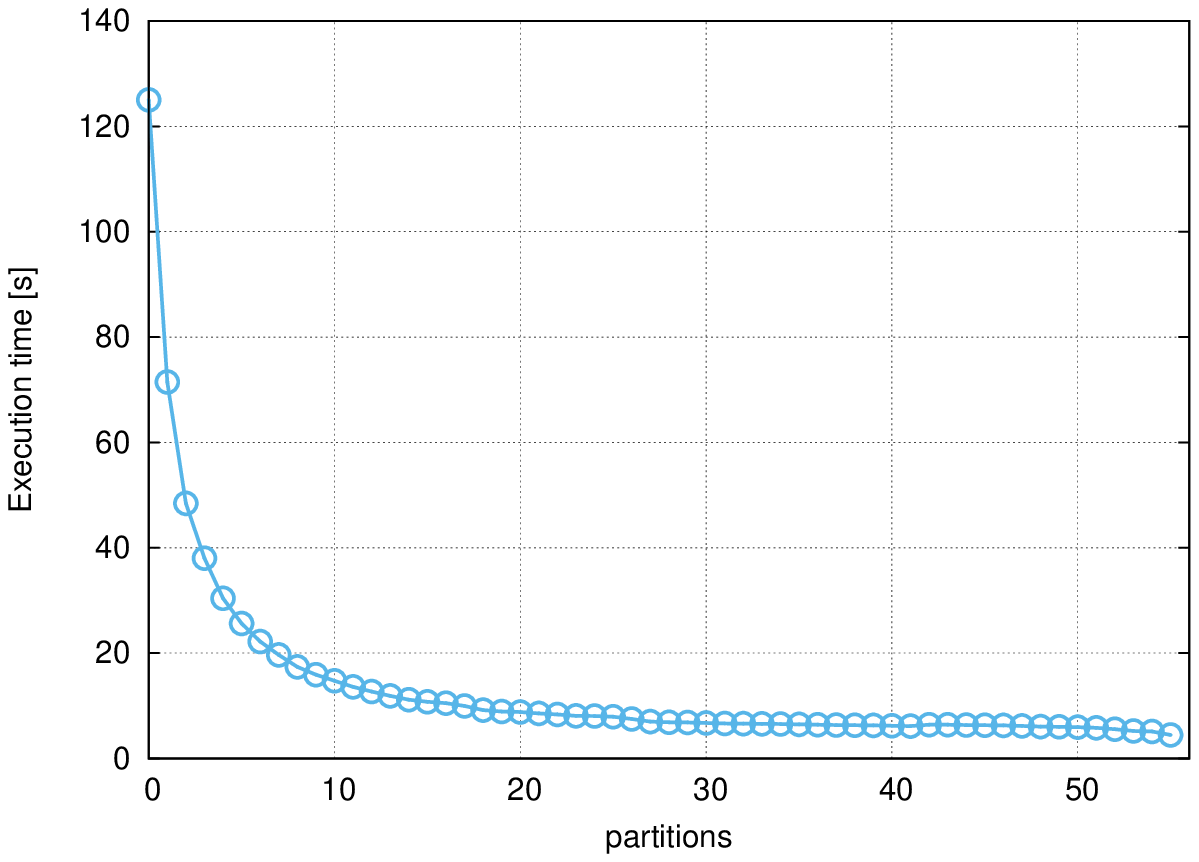}}
\subfigure[Hotspot.]{\label{fig:partitions:hs}\includegraphics[width=0.32\textwidth]{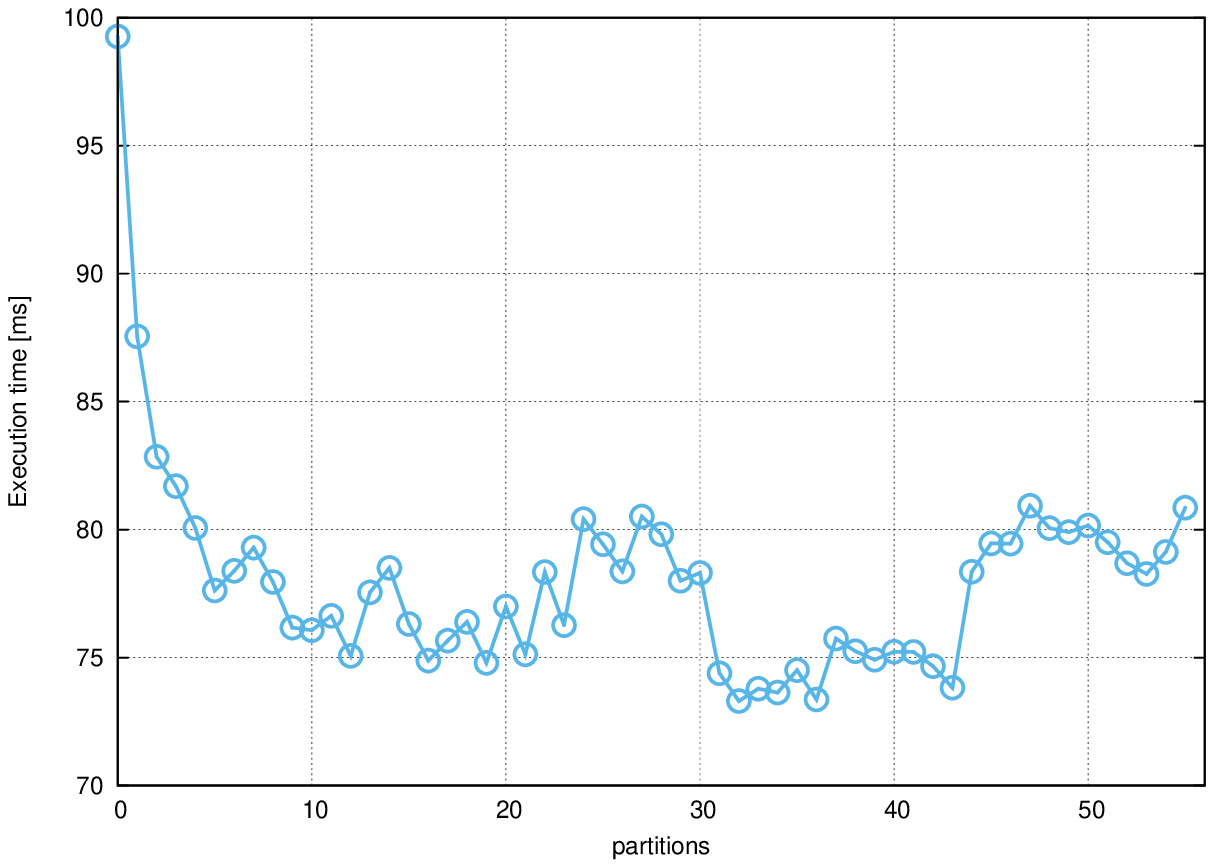}}
\subfigure[NN.]{\label{fig:partitions:nn}\includegraphics[width=0.32\textwidth]{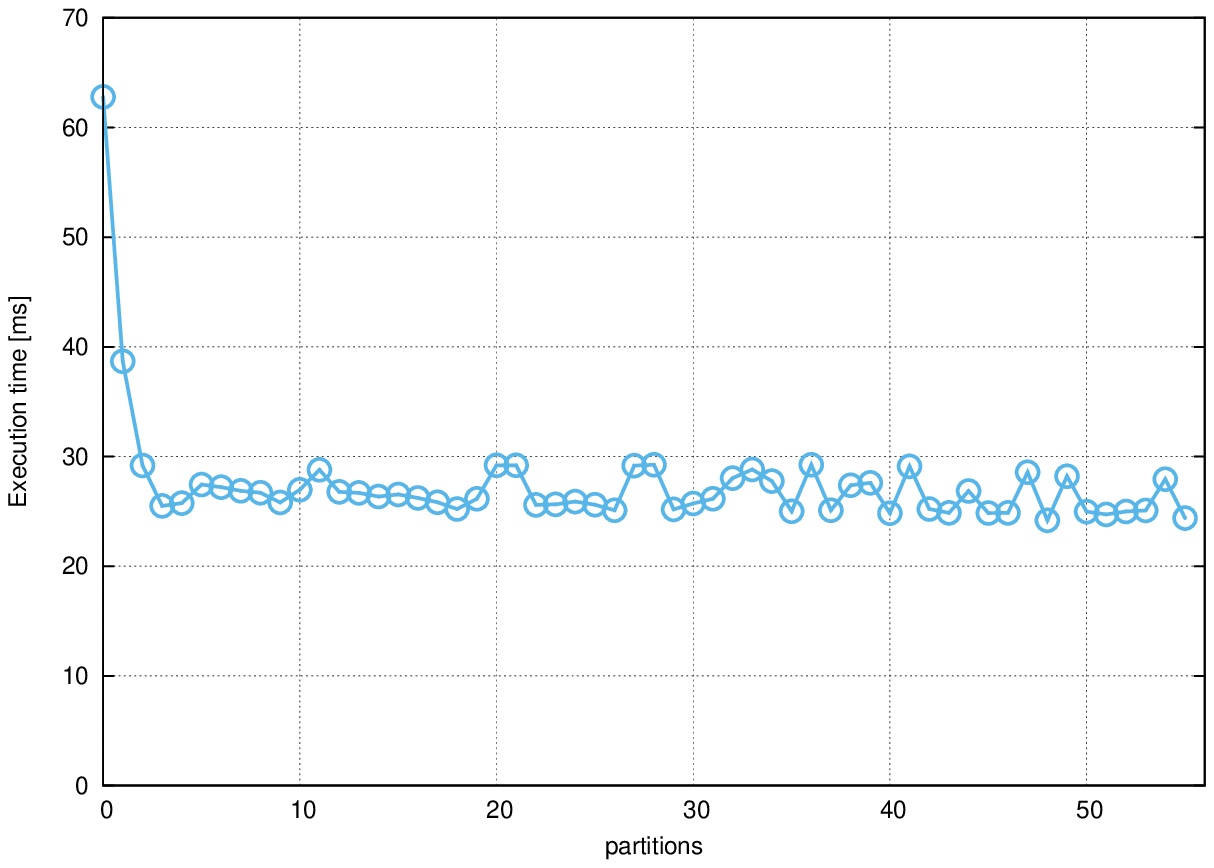}}
\subfigure[SRAD.]{\label{fig:partitions:srad}\includegraphics[width=0.32\textwidth]{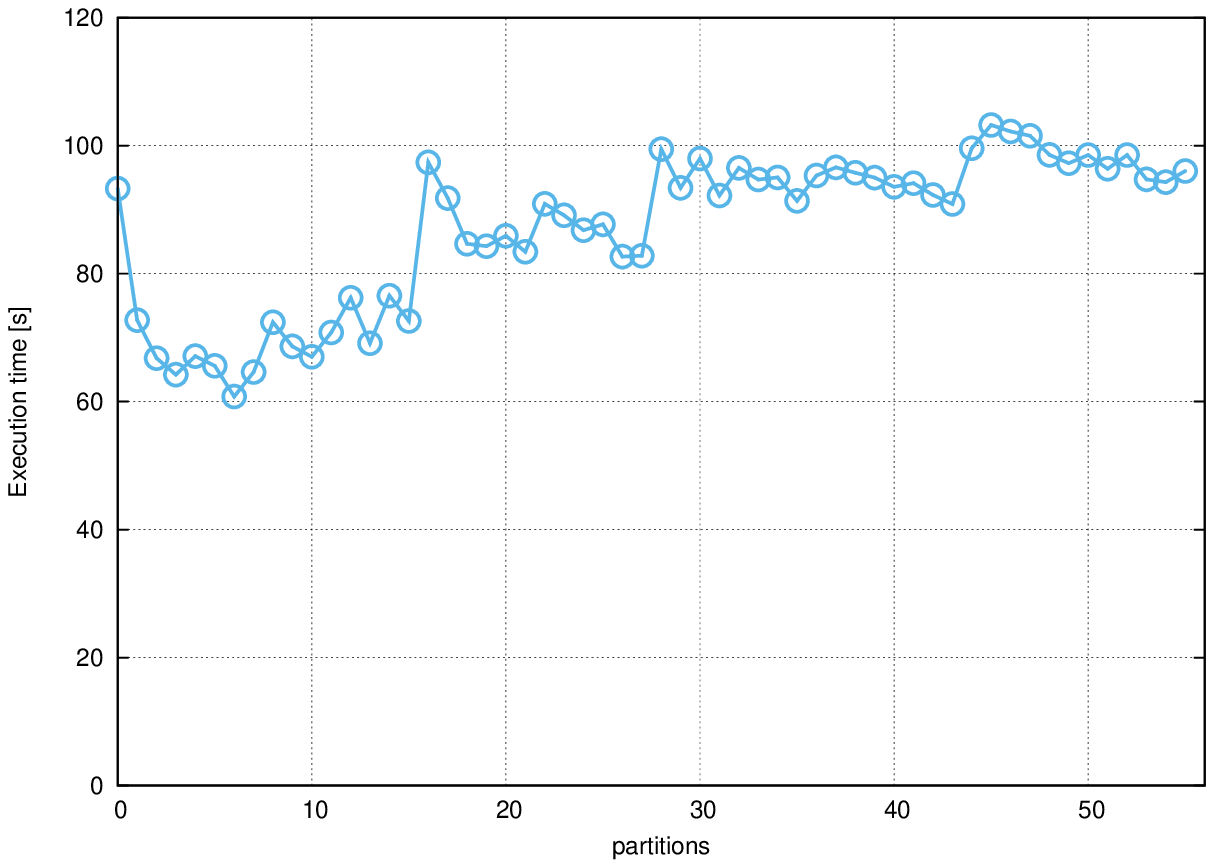}}
\caption{How the performance changes with the number of partitions. For Matrix Multiplication, $D=6000$, and $T=500 \times 500$; For Cholesky Factorization, $D=9600$, $T=800 \times 800$, and we use the \textit{column-major} layout. For Kmeans, $D=1120000$, $T=20000$, and we run 100 iterations before reaching convergence. For Hotspot, the grid size $D$ is $16384 \times 16384$, $T=1024 \times 1024$, and we run 50 iterations. For NN, the number of records $D$ is $5242880$, $T=512$, and the number of nearest neighbors to find is 10. For SRAD, $D=10000 \times 10000$, $T=20 \times 20$, $\lambda=0.5$, and we run the kernel for 100 iterations. }
\label{fig:partitions}
\end{figure*}

Figure~\ref{fig:partitions} shows how the overall performance changes with the number of partitions ($P$) when fixing task granularity ($T$). We observe that the performance varies significantly over $P$ for the six applications. For \texttt{MM} and \texttt{CF}, the benchmarks run much faster on some points than the others. On these points, 56 is a multiple of $P$, i.e., $P \in \{2, 4, 7, 8, 14, 28, 56\}$. Each 31SP Phi has 57 cores and one is reserved for the uOS. Thus, we have 224 ($56 \times 4$) available threads. When mapping $N$ streams onto a Phi, each stream will occupy $224/N$ threads. It is possible that two streams would share the same processing core and thus incur contention for shared resources such as caches. When the Phi is partitioned into groups, using these values within the set can avoid that the threads from the same core are partitioned into different streams. Therefore, we recommend using the number of partitions within the set $\{2, 4, 7, 8, 14, 28, 56\}$ for such applications.


For \texttt{Kmeans}, the execution time drops over the number of partitions (shown in Figure~\ref{fig:partitions:km}). When looking into the code, we observe \texttt{Kmeans} has to allocate and free temporal memory space dynamically in each iteration.  This overhead increases linearly with the number of threads and decreases with the number of streams accordingly. Thus, the performance trend of the non-overlappable \texttt{Kmeans} does not fit the one shown in Figure~\ref{fig:ubench:kernel}. 

For \texttt{Hotspot}, we see that the execution time roughly matches the trend shown in Figure~\ref{fig:ubench:kernel}. However, there are some fluctuations when changing the number of partitions. Particularly, when the number of partitions ranges from 33 to 37, the simulation time reaches its lowest points. At this time, the number of threads per partition is 6 or 7, and each partition will use threads on at most two processing cores. We believe this configuration will lead to a good cache utilization. 

Figure~\ref{fig:partitions:nn} shows how the number of partitions impacts the overall performance of \texttt{NN}. We see that the execution time first decreases sharply over partitions (and streams) til $P=4$. This is due to the fact that using more streams will create more opportunities of overlapping data transfers and kernel execution. Thereafter, the execution time remains around 25 ms. This overlappable application can use both temporal sharing and spatial sharing as shown in Figure~\ref{fig:flow:nn} of Section~\ref{subsec:benchmarks}. 

Figure~\ref{fig:partitions:srad} shows how the execution time changes over partitions for \texttt{SRAD}. On the whole, we notice that the performance first increases and then decreases, which roughly fits the trend presented in Figure~\ref{fig:ubench:kernel} of Section~\ref{subsec:ubench:kernel}. This is because this application consists of several kernels between which an explicit synchronization is needed. Thus, the application can only exploit spatial sharing of multiple streams. 


\subsubsection{How the number of tiles impacts performance?}  \label{subsec:titles}

\begin{figure*}[!t]
\centering
\subfigure[MM.]{\label{fig:tile:mm}\includegraphics[width=0.32\textwidth]{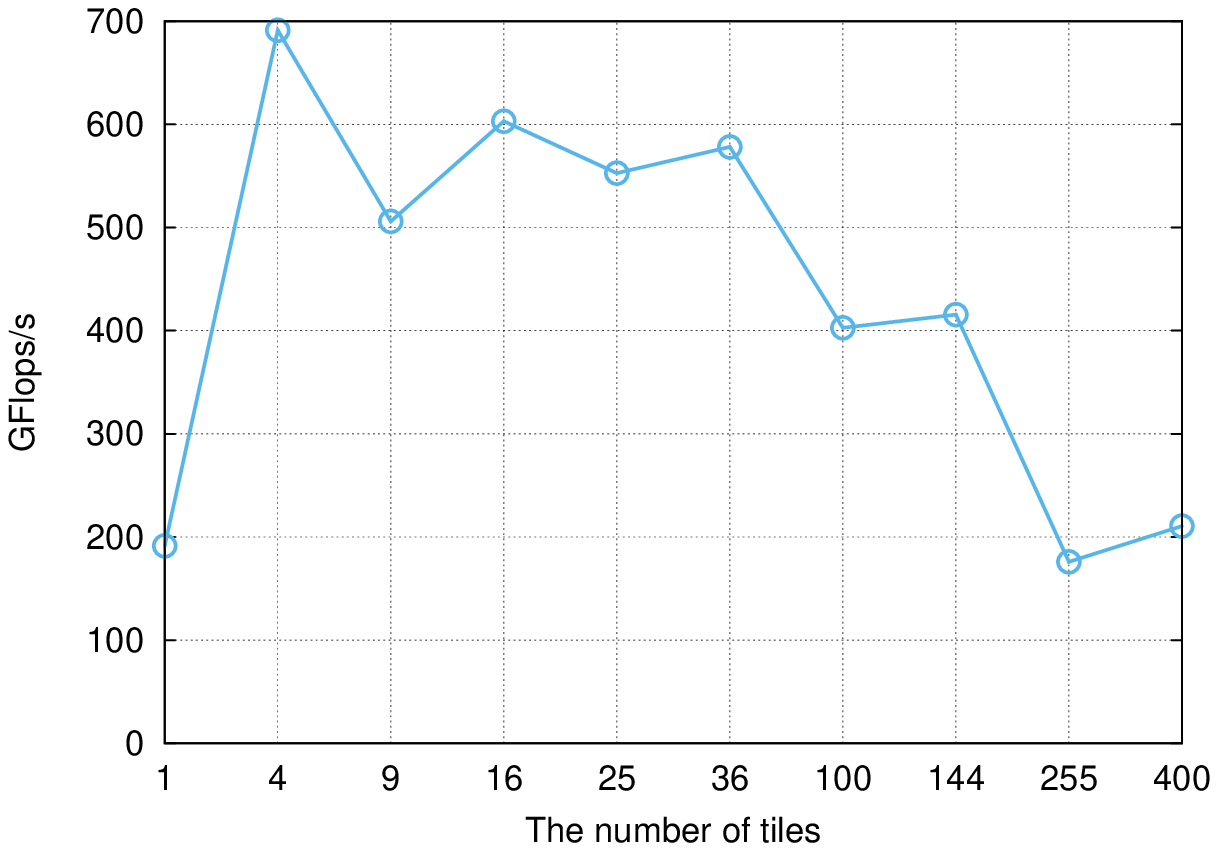}}
\subfigure[CF.]{\label{fig:tile:cf}\includegraphics[width=0.32\textwidth]{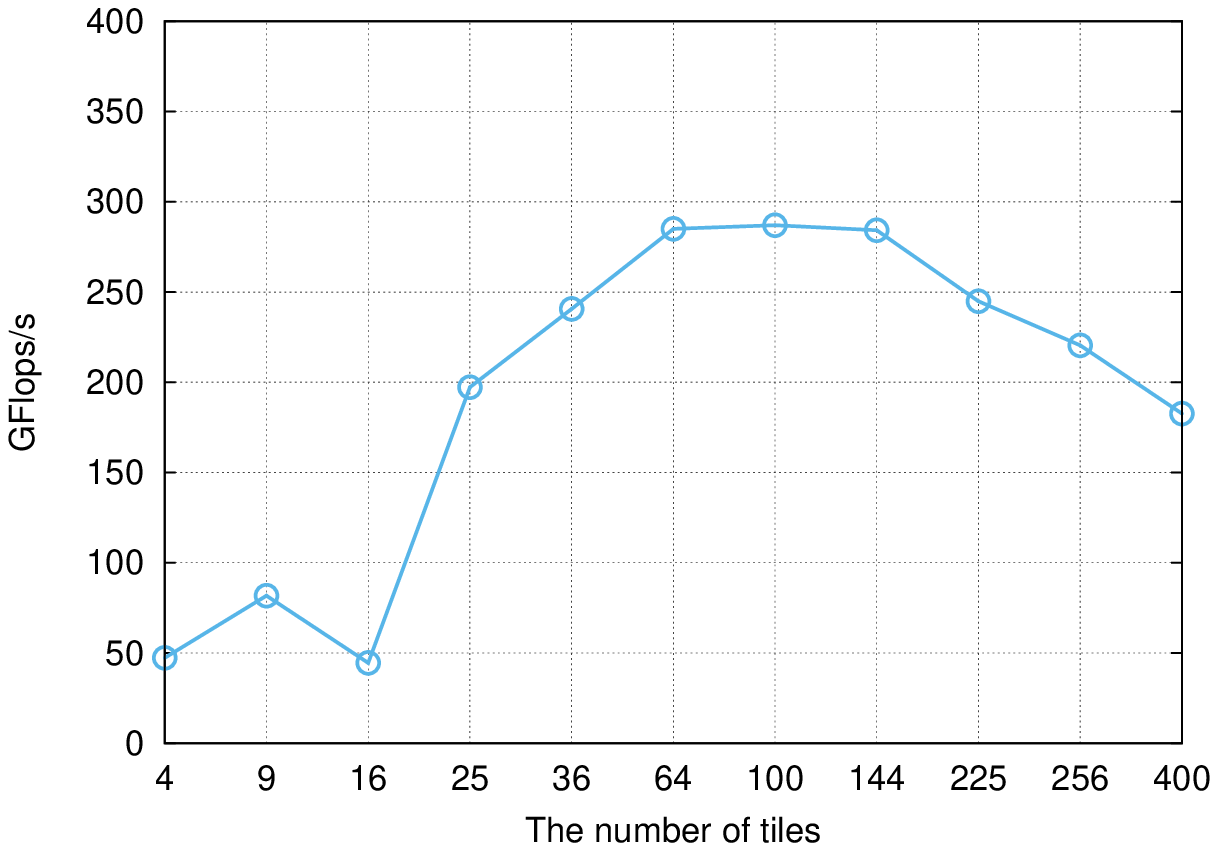}}
\subfigure[Kmeans.]{\label{fig:tile:km}\includegraphics[width=0.32\textwidth]{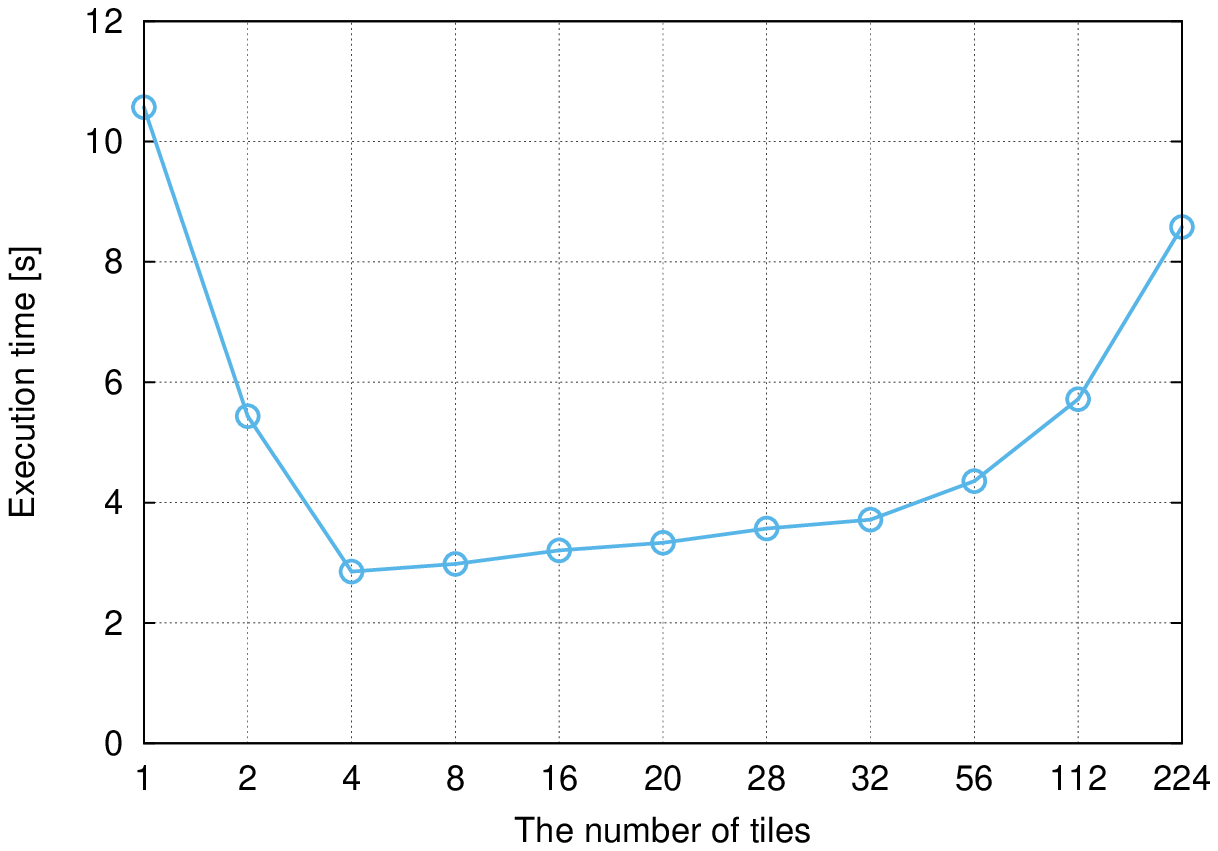}}
\subfigure[Hotspot.]{\label{fig:tile:hs}\includegraphics[width=0.32\textwidth]{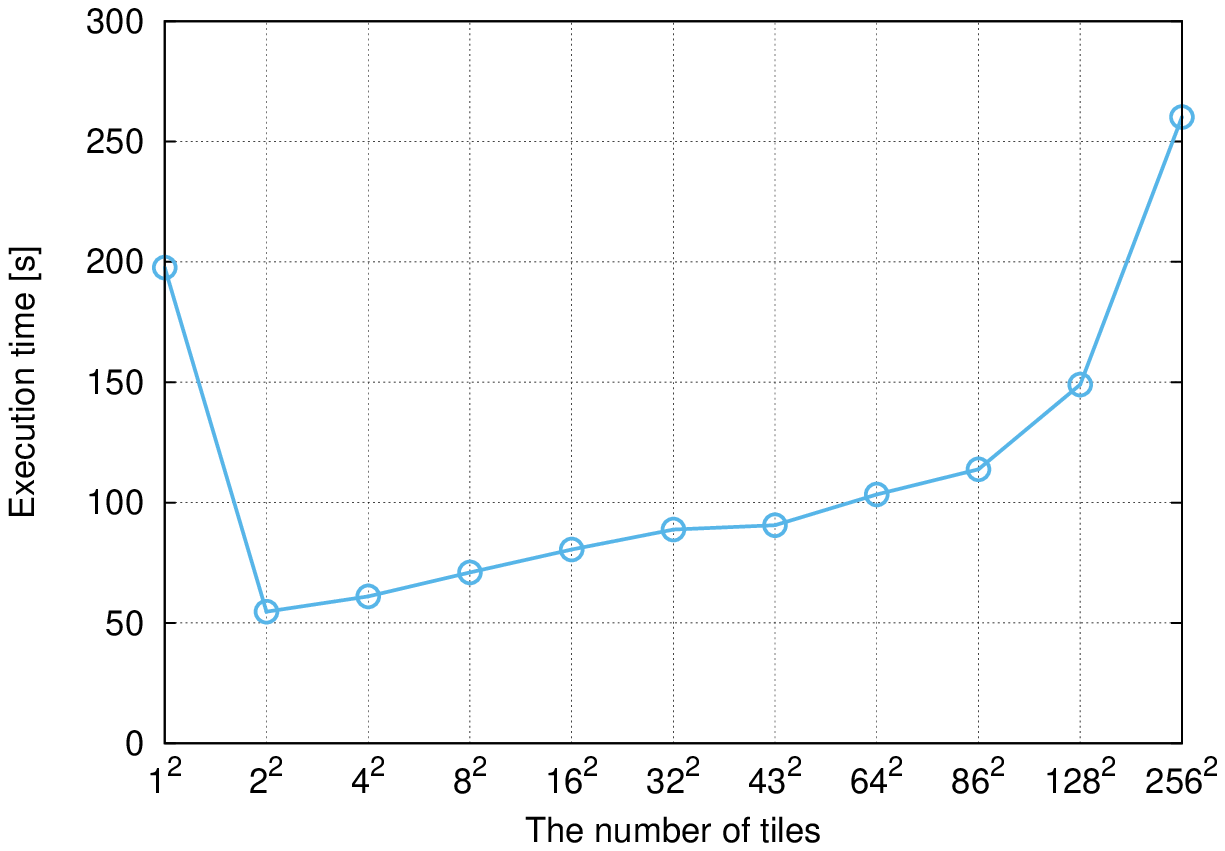}}
\subfigure[NN.]{\label{fig:tile:nn}\includegraphics[width=0.32\textwidth]{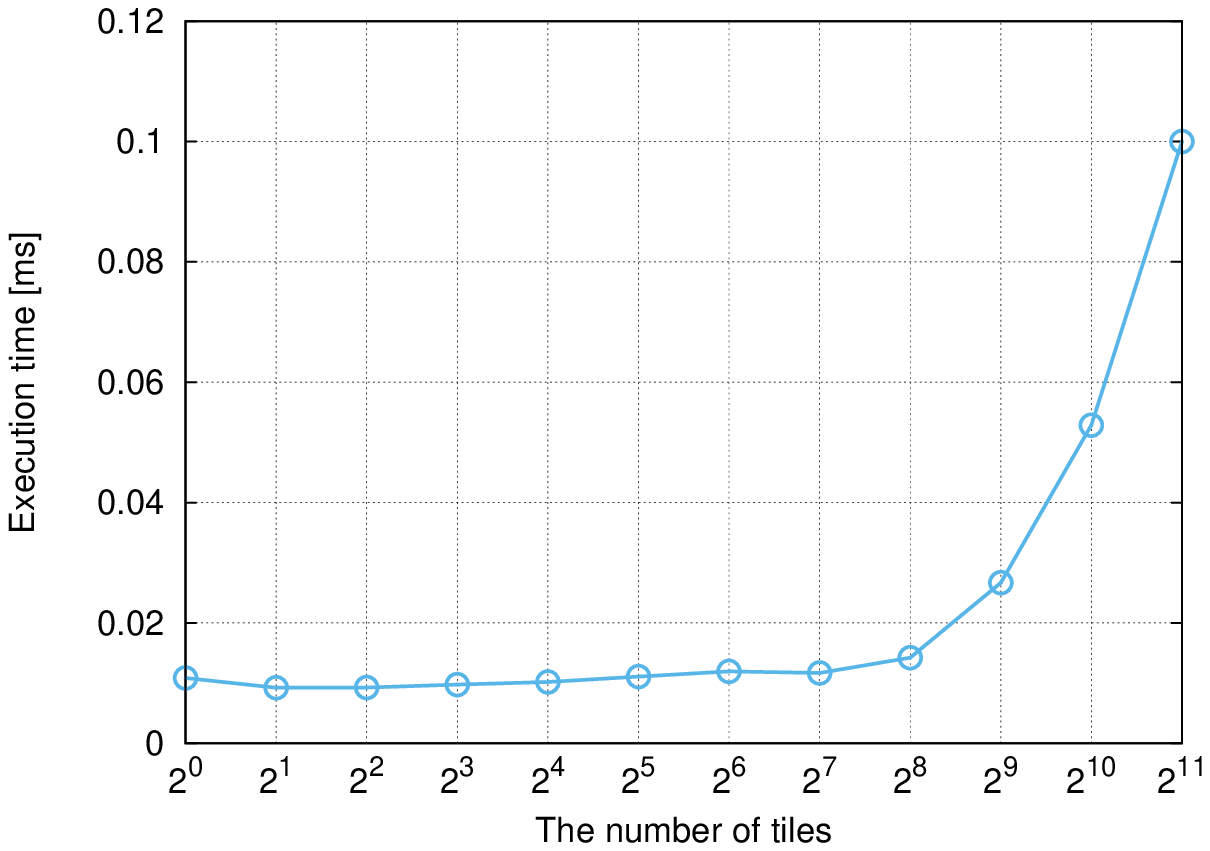}}
\subfigure[SRAD.]{\label{fig:tile:srad}\includegraphics[width=0.32\textwidth]{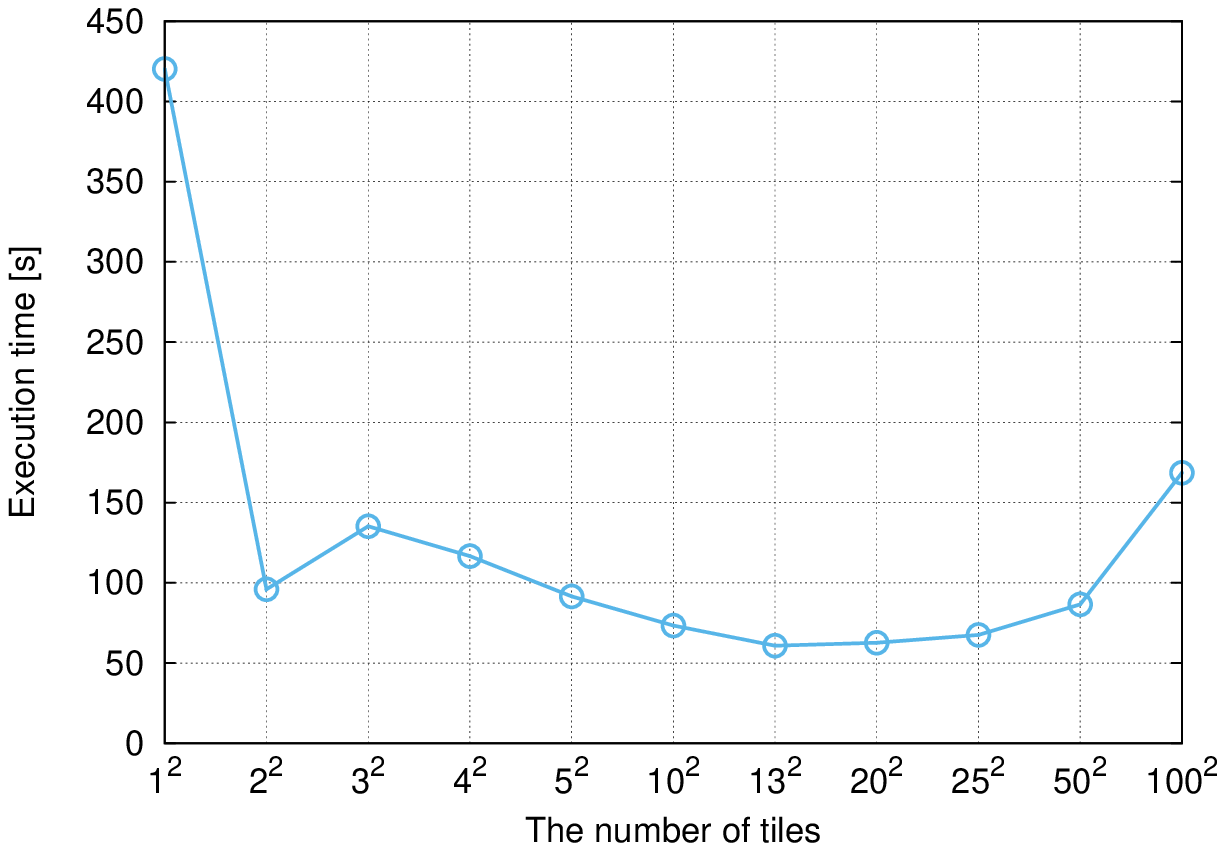}}
\caption{How the performance changes with the number of tiles. For Matrix Multiplication, $D=6000$, and $P=4$; For Cholesky Factorization, $D=9600$, $P=4$, and we use the \textit{column-major} layout. For Kmeans, $D=1120000$, $P=4$, the number of centroid is 8, and we run 100 iterations before reaching convergence. For Hotspot, the grid size $D$ is $16384 \times 16384$, $P=4$, and we run 50 iterations. For NN, the number of records $D$ is $5242880$, $P=512$, the target coordination is (40, 120), and the number of nearest neighbors to find is 10. For SRAD, $D=10000 \times 10000$, $P=4$, $\lambda=0.5$, and we run the kernel for 100 iterations. }
\label{fig:tile}
\end{figure*}

Figure~\ref{fig:tile} shows how the performance changes with the number of tiles ($T$) for the six applications. Overall, the achieved performance first increases and then decreases (note the different metrics between \texttt{MM}, \texttt{CF} and the other four applications). In particular, we observe most applications run the fastest when the number of tiles/tasks is 4. With one task, the performance decreases sharply. This is due to the fact that we partition the 56 cores of a Phi into four groups ($P=4$), and mapping the tile to a partition will leave the other partitions idle. Further, using a larger $T$ (i.e., more tiles but each tile is smaller) will have more pipelining opportunities to overlap stalls. But using a large $T$ introduces extra control overheads and incurs a relatively low resource utilization. Therefore, further increasing the number of tasks leads to a worse performance as shown in Figure~\ref{fig:tile}. 


In addition, selecting $T$ for \texttt{CF} and \texttt{SRAD} differs from the other applications. Their achieved performance reaches the optimal when $T=100$, and $T=400$, respectively. Different from other applications, these two contain several kernels which could introduce context interference. Furthermore, we see that \texttt{NN} obtains a similar performance between $T=1$ and $T=4$. This is because \texttt{NN}'s performance is bounded by data transfers and creating multiple streams to achieve overlapping brings a slight difference to the overall performance.

\subsection{Discussion} \label{sec:discussion}

\subsubsection{Using Occasion}


Our experimental results show that using multiple streams is beneficial only when the applications are \textit{overlappable}. For such applications, exploiting temporal sharing of hardware resources will overlap data transfers and kernel execution, and thus speedup the execution process. Further, using spatial sharing of hardware resources will increase the resource utilization. Note that only leveraging spatial sharing might not lead to a performance improvement for the non-overlappable applications. 

Moreover, we observe several special cases that using multiple streams is beneficial for the non-overlappable applications (e.g., \texttt{Kmeans} and \texttt{SRAD}). This is because of the extra kernel overheads, e.g., allocating/deallocating temporal memroy space. Therefore, we have to consider such application-specific characteristics when using multiple streams.

\subsubsection{Reducing the Search Space} \label{sec:perf_model}
As can be seen from Section~\ref{subsec:perf:analysis}, resource granularity ($P$) and task granularity ($T$) have a significant impact on the overall performance. For a given application, maximizing the overall performance need search for the optimal value for each factor. This will consume a huge amount of time. Hereby we discuss how to prune the search space when selecting a proper value for $P$ and $T$.


As indicated in Figure~\ref{fig:partitions:mm} and Figure~\ref{fig:partitions:cf}, we obtain that $P \in \{2, 4, 7, 8, 14, 28, 56\}$ and such values will avoid that the threads from the same core are mapped to different streams. The results of the other overlappable application (\texttt{NN}) show that when $P \ge 4$, the performance remains around 25 ms. Therefore, we should focus our attention on these special numbers. 


When determining the number of tiles, the first priority is to guarantee load balancing. This is particularly true when $T<P$, i.e., the resource is under-utilized (Figure~\ref{fig:tile}). Therefore, we guarantee that $T=m \cdot P$, where $m \in \{1, 2, 3, ...\}$.  Besides, $T$ should not be too large to achieve a good resource utilization, and it should not be too small to exploit the pipelining potentials. 







To summarize, to achieve the optimal performance for will incur a huge search space. Our guidelines reduce the search space significantly. To further reduce the search space, we need a fine analytical performance model~\cite{citeulike:9715521}\cite{citeulike:13920334}\cite{citeulike:13920353}. Alternatively, we plan to use machine learning techniques to obtain a proper value for $P$ and $T$.


  	
\section{Preliminary Results on Multiple MICs} \label{sec:multimics}
Current large-scale computing systems often employ multiple accelerators to guarantee its peak performance. Thus, how to use multiple devices simultaneously becomes an issue. Using multiple streams seems a promising tool. For example, \texttt{hStreams} provides a unified resource management layer of all the Phis (MICs) and its runtime automatically map the streams to the underlying hardware domain. In this way, a streamed code can run on multiple Phis without code modifications. In this section, we discuss the preliminary results on the heterogeneous platforms with multiple Phis. 

\begin{figure}[!t]
\centering
\includegraphics[width=0.40\textwidth]{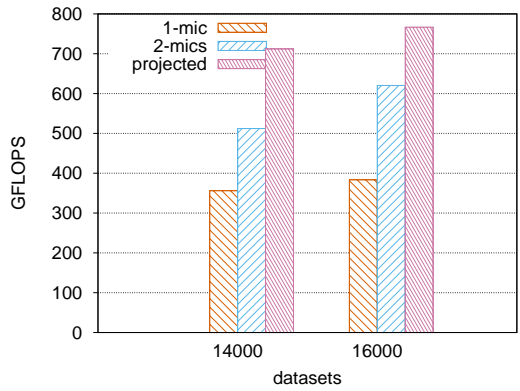}
\caption{How Cholesky Factorization performs on multiple MICs for datasets $14000 \times 14000$ and $16000 \times 16000$.} 
\label{fig:multimics}
\end{figure}

Figure~\ref{fig:multimics} shows the \texttt{CF}'s performance on one and two Phis. We see that the achieved performance increases significantly with two devices than with one device, but the performance is still lower than the projected performance of two Phis. This is because partitioning workloads among devices with separate memory space needs to transfer more data blocks than that using only one device. Also, \texttt{CF} contains several kernels and explicit synchronizations are required between them. When using multiple Phis, synchronizations between streams from different Phis might introduce extra overheads. To gain more insights, we would like to run more experiments with a wide range of applications in future. 

\section{Conclusion} \label{sec:conclusion}

The potential of using multiple streams on the heterogeneous platforms is expected to be significant for a wide range of applications. In this paper, we perform a systematic performance evaluation of multiple streams on the MIC-based heterogeneous platforms. Our experimental results at the microbenchmarking level and the real-world application level lead to the following observations/conclusions: (1) The data transfers in both directions on Phi cannot run concurrently; (2) Data transferring on Phi overlaps kernel execution, but the full overlap seems not achievable; (3) Using multiple streams might not lead to a performance increase only in the presence of spatial resource sharing; (4) Being overlappable is a must for benefits when using multiple streams; (5) Both task granularity and resource granularity have a large impact on the overall performance; (6) Some non-overlappable application still enjoy a performance improvement by using multiple streams. 

In the future, we would like to investigate how to transform the non-overlappable applications to overlappable applications. Further, we will leverage machine learning techniques to obtain a proper task and resource granularity. Also, we plan to further evaluate the performance impact on multiple Phis.

\section*{Acknowledgment}

We would like to thank the authors from the Rodinia benchmark suite for their valuable benchmarks. We are also thankful to the reviewers for their constructive comments. This work was partially funded by the National Natural Science Foundation of China under Grant No.61402488 and No.61502514, the National High-tech R\&D Program of China (863 Program) under Grant No. 2015AA01A301, the National Research Fundation for the Doctoral Program of Higher Education of China (RFDP) under Grant No. 20134307120035 and No. 20134307120031. 

\bibliography{mybib}{}
\bibliographystyle{ieeetr}

\end{document}